\documentclass[10pt,conference]{IEEEtran}

\usepackage{hyperref}
\usepackage{xspace}

\usepackage{soul}

\usepackage{flushend}

\newcommand{\citet}[1]{\textcite{#1}}

\usepackage[T1]{fontenc}
\usepackage{url}
\usepackage{amsmath}
\usepackage{colortbl}
\usepackage{ctable}
\usepackage{enumitem}

\usepackage[caption=false]{subfig}
\usepackage{multirow}
\usepackage{adjustbox}
\usepackage{rotating}
\graphicspath{{figures/}}
\usepackage{graphicx}

\usepackage{listings}

\makeatletter
\lst@Key{countblanklines}{true}[t]%
    {\lstKV@SetIf{#1}\lst@ifcountblanklines}

\lst@AddToHook{OnEmptyLine}{%
    \lst@ifnumberblanklines\else%
       \lst@ifcountblanklines\else%
         \advance\c@lstnumber-\@ne\relax%
       \fi%
    \fi}
\makeatother

\definecolor{codeblack}{rgb}{0,0,0}
\definecolor{lightgray}{gray}{0.98}
\definecolor{codeblue}{rgb}{0.13,0.13,1}
\definecolor{codegrey}{rgb}{0.36,0.35,0.38}
\definecolor{codegreen}{rgb}{0,0.5,0}
\definecolor{codered}{rgb}{0.9,0,0}
\definecolor{OliveGreen}{cmyk}{0.64,0,0.95,0.40}
\definecolor{DarkGreen}{cmyk}{0.40,0,0.50,0.15}
\definecolor{purple}{cmyk}{0.41,0.73,0,0.1}
\definecolor{LightRed}{cmyk}{0,0.682,0.728,0.6}

\lstdefinestyle{default}{
    escapechar=\$,
	backgroundcolor=\color{lightgray},
	keywordstyle=[1]\color{codeblue},
	keywordstyle=[2]\color{purple},
	keywordstyle=[3]\color{codered},
 	stringstyle=\color{codegreen},
	commentstyle=\color{codegrey},
	basicstyle=\color{codeblack}\footnotesize\ttfamily,
	columns=fullflexible,
    breaklines=true,
	prebreak=\raisebox{0ex}[0ex][0ex]{\ensuremath{\hookleftarrow}},
    xleftmargin=6pt,
    xrightmargin=0pt, 
	frame=none,
    framexleftmargin=0pt,
    framexrightmargin=0pt,
    mathescape=true,
	numbers=left,
	numberblanklines=false,
    countblanklines=false,
	stepnumber=1,
	numbersep=4pt,
	numberstyle=\footnotesize,
	keepspaces=true,
    showspaces=false,
    showstringspaces=false,
	showtabs=false,
	upquote=true,
    tabsize=2
}

\lstdefinestyle{boa}{
	mathescape=false,
	escapechar=\%,
	emphstyle=\color{purple},
	emphstyle={[2]\color{LightRed}},
	morestring=[b]`,
	gobble=0,
	morekeywords={_,argument,input,view,exists,foreach,ifall,output,of,weight,stop,visit,before,after,switch,default,current,view},
	emph={int,string,bool,table,float,time,array,stack,map,visitor,%
true,false,%
top,sum,mean,maximum,minimum,set,collection,bottom,%
Project,Person,CodeRepository,Revision,ChangedFile,ASTRoot,Namespace,Declaration,Type,Method,Variable,Statement,Expression,Modifier,IfStatement,%
BooleanLiteral,StringLiteral,PDG,%
InfixExpr,PrefixExpr,PostfixExpr,ForStatement,DoStatement,WhileStatement,TryStatement,ReturnStatement,ThrowStatement,ContinueStatement,BreakStatement,LabeledStatement,%
ExpressionKind,NEW,LITERAL,EQ,NEQ,OP_ADD,OP_MULT,OP_SUB,OP_INC,OP_DEC,BIT_NOT,LOGICAL_NOT,%
TypeKind,CLASS,ANONYMOUS,%
ModifierKind,OTHER,%
ChangeKind,DELETED,%
StatementKind,IF,BREAK,RETURN,THROW,CONTINUE,LABEL,FOR,DO,WHILE,SWITCH,TRY,%
RepositoryKind,SVN},
	emph={[2]logDead,isboollit,isstringlit,isinfix,isprefix,ispostfix,isfixingrevision,getast,iskind,hasfiletype,isliteral,getsnapshot,has_modifier_public,%
new,clear,values,contains,keys,add,format,def,len,match,lowercase,yearof,haskey,remove,strfind,push,pop,peek,normalize,getpdg,getcrypthash,gettotalnodes,gettotaledges,gettotalcontrolnodes,getpdgslice,_row,chainlen},
}

\lstnewenvironment{Boa}
  {\lstset{language=bash,style=default,style=boa}}
  {}

\lstnewenvironment{Java}
  {\lstset{language=Java,style=default}}
  {}

\newcommand{\figref}[1]{Figure~\ref{#1}}
\newcommand{\secref}[1]{Section~\ref{#1}}
\newcommand{\tabref}[1]{Table~\ref{#1}}
\newcommand{\lineref}[1]{line~\ref*{#1}}
\newcommand{\linesref}[2]{lines~\ref*{#1}--\ref*{#2}}

\usepackage{mdframed}
\usepackage{xpatch}

\makeatletter
\xpatchcmd{\endmdframed}
  {\aftergroup\endmdf@trivlist\color@endgroup}
  {\endmdf@trivlist\color@endgroup\@doendpe}
  {}{}
\makeatother
\newcounter{FindingCounter}
\stepcounter{FindingCounter}
\newcommand{\findings}[1]{
	\begin{mdframed}[backgroundcolor=gray!10,
			linewidth=0.75pt,
			roundcorner=5pt,
			innertopmargin=2mm,
			innerbottommargin=2mm,
			innerrightmargin=2mm,
			innerleftmargin=2mm,
			skipabove=1mm,
			skipbelow=1mm,
			font=\small]
		{\bf{Finding~\arabic{FindingCounter}:}}~ #1
	\end{mdframed}
	\stepcounter{FindingCounter}
}

\usepackage[style=ieee,mincitenames=1,maxcitenames=1,sorting=nty]{biblatex}
\usepackage{balance}

\begin{document}

\title{Method Chaining Redux: An Empirical Study of Method Chaining in Java, Kotlin, and Python}

\author{
    \IEEEauthorblockN{Ali M. Keshk}
    \IEEEauthorblockA{University of Nebraska--Lincoln\\
        Email: \url{akeshk2@huskers.unl.edu}}
        \and
    \IEEEauthorblockN{Robert Dyer}
    \IEEEauthorblockA{University of Nebraska--Lincoln\\
        Email: \url{rdyer@unl.edu}}
}

\maketitle
\thispagestyle{plain}
\pagestyle{plain}

\begin{abstract}
There are possible benefits and drawbacks to chaining methods together, as is often done in fluent APIs. A prior study investigated how Java developers chain methods in over 2.7k open-source projects. That study observed, for the dataset analyzed, that the use of method chaining in Java is popular and seems to be increasing over time. That study however was limited to a smaller sample of Java projects, and it is also not clear if the results generalize to other languages. In this work, we first replicate the prior results by building a similar dataset and our own analysis scripts. We then extend those results by analyzing a much larger dataset of 89k Java projects and generalizing to other programming languages by analyzing 26k Kotlin projects and 98k Python projects. The results show chaining is more popular in Java and Kotlin than Python, chaining use in Kotlin is not growing, and Python sees more use in non-testing code.
\end{abstract}

\begin{IEEEkeywords}
method chaining, empirical study, replication, Java, Kotlin, Python
\end{IEEEkeywords}

\section{Introduction}
\label{sec:intro}

Most object-oriented languages support \textit{chaining} method calls together, to avoid the need to temporarily store a returned object the programmer intends to only immediately use as the receiver for the next method call.  Any arbitrary number of method calls can be chained together and the receiver object used can either be the same for every call, or possibly different (if one call returns another object).  For example, the following code builds a JSON string by chaining several methods together then calling \lstinline|toString()|:

\begin{Java}
String json = new JSONObject()
                 .put("Conference", "MSR")
                 .put("Year", "2023")
                 .toString();
\end{Java}

Method chaining provides several potential benefits, including: elimination of storing the temporary objects returned~\cite{Fowler10}, improved readability of internal domain-specific languages (DSL)~\cite{Fowler10}, the ability to more easily skip optional arguments in methods with many arguments~\cite{lukas10}, and DSLs can build expressions that read more natural from left to right~\cite{Fowler10,linq}.  Method chaining is also commonly incorporated in fluent APIs~\cite{fluent}, and is often used in design patterns such as the builder pattern~\cite{gamma1994design} (shown above).

Despite the apparent usefulness of the approach, there has been a lot of discussion about it.  Some people claim the use of method chaining is generally a bad practice~\cite{so-conflict}, possibly causing readability/comprehension problems~\cite{Kasraee946689,borstler16,scheller2013usability}, making it more difficult to debug in some debuggers, or even by breaking the Law of Demeter~\cite{demeter,borstler16}.  Others claim method chaining leads to maintenance issues~\cite{fluentbad}.

Knowledge of how developers use method chaining could benefit the research community by providing crucial evidence to lay the groundwork for future studies.  For example, knowing if developers employ method chaining could help guide researchers interested in the reasons to use chaining.  Does the type system (static vs. dynamic) play a role?  Or maybe the language does not provide fluent style APIs?

Thus, knowing if method chaining is utilized by the programming community or not could be important for language designers.  Such knowledge could help guide future language and API/library designs.  If method chaining is popular in a language, the maintainers could optimize their compiler or virtual machine to account for method chains.

This study could also directly benefit practitioners, especially those that write APIs, as they could be made aware of which language(s) users seem to employ method chaining in.  E.g., if someone is writing a new Python API, they might consider a non-fluent design if Python developers shy away from chaining, whereas the Kotlin API might decide to be more fluent if Kotlin developers utilize chaining.

Since it was not entirely clear if the programming community accepts the idea of method chaining or not, \citet{original-study} studied how method chaining was (or was not) adopted in the Java programming language.  They performed an empirical study on over 2.7k popular open-source Java projects from GitHub and looked at how often method chains were used.  Their results indicated that method chaining is relatively popular among Java projects and that the use of method chaining was increasing over time.  They also observed method chaining was more popular in testing files (vs non-testing files), and they proposed a set of language enhancements to Java to encourage additional uses of method chaining.  But their study focused only on a single programming language.

For example, one of the perceived benefits of method chaining is the ability to more easily skip over optional arguments of a method with many arguments.  In Java for example, if a method takes 10 arguments but many are optional, the API designer has to either provide a large amount of similar looking overrides, or fall back to a fluent API design.  Here, the fluent design using method chains would be more flexible and (most likely) easier to comprehend.

But this is only a benefit in certain languages.  For example, Kotlin and Python allow specifying default values for arguments and calling methods with named/keyword arguments and so do not have the same difficulty with skipping optional arguments as one might have in a language like Java.  Thus it is important to understand how the use of method chaining might differ in other languages that provide additional features that might discourage the need to chain methods.

In this work, we first investigate if it is possible to replicate the results of \citet{original-study} by building a similar dataset and writing our own analysis scripts.  We built a dataset containing  projects from their dataset (but cloned several years later) and discovered a few small inconsistencies when compared to their results.  We confirmed with the original authors these were bugs in the analysis and our results indicate those bugs did not change the overall previous results.

We then extend the study to a much larger dataset of Java projects, with 89k (35x more) projects, to see if the results generalize to a larger population.  Our results show they do: we observe similar trends in the larger dataset as the original study observed in their smaller sample.  What was not clear was how the results generalize to other languages.

To investigate other languages, we also used datasets with 26k Kotlin projects and 98k Python projects.  We chose Kotlin as it is one of the top-ten statically-typed languages~\cite{tiobe}, the default language for Android, and designed to interoperate well with Java.  Thus we wondered if the trends might be similar to Java, despite any language differences, as many of the developers are also Java developers.  We chose Python as it is a popular scripting language with some syntax and style differences like enforcing whitespace (developers need to either enclose the whole chain in parentheses or end each line with a backslash) that could affect the use of method chaining.  Thus we suspected Python developers might behave differently compared to Java/Kotlin developers.

The results show chaining is more popular in Java and Kotlin than Python, its use is not growing in Kotlin, and Python sees more use in non-tests than testing code.  Given the prevalence of chaining in Java and Kotlin, practitioners need to be made aware of what method chaining is, how best to utilize it, and what design patterns are built on top of it.  Even if they themselves are not writing code using method chains, they are very likely to stumble across such code.  We need to ensure new developers are properly trained and there is sufficient documentation to support them.  Conversely, it seems Python developers could possibly avoid these issues, as chaining is almost three times less prevalent.  These results also give evidence that API developers can feel comfortable utilizing fluent designs when targeting Java or Kotlin, while Python library developers may want to avoid a fluent design.

In the next section, we give background information on method chains and the prior study.  In \secref{sec:rqs}, we discuss this study's research questions.  The approach is overviewed in \secref{sec:approach} and results provided in \secref{sec:results}.  In \secref{sec:threats}, we discuss threats to the validity of the study.  Related work is discussed in \secref{sec:related}, and we conclude in \secref{sec:conclusion}.

\section{Background}
\label{sec:background}

In this section we give background defining what a method chain is along with some clarifying examples.  Then, we summarize the prior study and its main findings.

\subsection{Method Chains and Chain Length}

A \textbf{method invocation} issues a call to a method and requires a receiver object, hence constructor calls or \lstinline|super()| calls inside a constructor are not method chains.  Similar to \citet{original-study} (that we also refer to as the ``original study'' throughout this paper), we define a \textbf{method chain} as ``a sequence of one or more method invocations joined by the `.' symbol''~\cite{original-study} and also define the \textbf{length} of a method chain ``as the number of [method] invocations in the sequence''~\cite{original-study}.

\begin{figure}[ht]
\begin{Java}
 new C();     // not a method chain$\label{ln:ex1}$
 super();     // not a method chain$\label{ln:ex2}$

 // length: 1
 o.m();       // explicit receiver$\label{ln:ex3}$
 m();         // implicit 'this' receiver$\label{ln:ex4}$
 new C().m(); // constructor not included$\label{ln:ex5}$
 super.m();   // super is not a call here$\label{ln:ex6}$

 // length: 2
 o.m().n();$\label{ln:ex7}$

 // two chains, each length: 1
 m(n());$\label{ln:ex8}$
 m().f.n();$\label{ln:ex9}$
\end{Java}
  \vspace{-0.5em}
    \caption{Example method chains in Java and their lengths}
    \label{fig:examples}
\end{figure}

In \figref{fig:examples} we show some example method chains in Java.  The first two examples simply show that, despite their appearance, constructor calls (\lineref{ln:ex1}) and calls to super constructors (\lineref{ln:ex2}) are not method chains.  The examples on \linesref{ln:ex3}{ln:ex6} all show single method invocations (which we call chain length 1), on varying receivers.  The example on \lineref{ln:ex7} is the first example of what most call a method chain, with length 2.  Some trickier cases are shown in \linesref{ln:ex8}{ln:ex9} where there can be more than one chain, if they are nested in the arguments of a call or separated by a field access.

\subsection{Prior Study}

The original study looked at popular Java projects from GitHub, discovered in Nov/Dec of 2019.  In total, they analyzed 2,756 Java projects and focused on the years 2010--2018, as they wanted full years of data (so they could not keep 2019) and they wanted enough files/projects, so started with 2010.  They then took repository snapshots for each year studied.

\vspace{0.5em}
\noindent\begin{minipage}{.4\linewidth}
\begin{equation}\label{eq:fn}
f_n = \dfrac{m_n}{m_1}
\end{equation}
\end{minipage}%
\begin{minipage}{.6\linewidth}
\begin{equation}\label{eq:r}
r = \dfrac{\sum_{n \ge 2}{n \cdot m_n}}{\sum_{n \ge 1}{n \cdot m_n}}
\end{equation}
\end{minipage}
\vspace{0.2em}

Their analysis relied on computing two metrics: $f_n$ and $r$.  $f_n$ (Equation~\ref{eq:fn}) is the number of chains of length $n$ over the number of not-chained method invocations (aka length 1)~\cite{original-study}, where $m_n$ is the number of chains of length $n$.  $f_n$ is not an average, but instead is computed for each file in the dataset, on a per-year basis.

The second metric they computed were the $r$ values (Equation~\ref{eq:r}), the ratio of all chained method invocations to all method invocations (chained or not)~\cite{original-study}, where this ratio is computed either per-project or over all files in the whole dataset, on a per-year basis.

The third metric was $U_n$: the ratio of projects containing at least one chain whose length is longer than or equal to $n$~\cite{original-study}.

At a high level, their results showed the use of method chaining in Java increased from 2010 to 2018.  The percentage of all method chains rose from 16.0\% in 2010 to 23.1\% in 2018.  They also found that over half of the Java projects contained chains of length $n \ge 8$, and less than 5\% of projects contained chains of length $n \ge 42$.  When looking at extra-long chains, the three most common libraries were: Elasticsearch, Guava, and the Java standard library.  They concluded that method chaining is most likely an accepted practice in Java due to the observed high and increasing number of uses.

\section{Research Questions}
\label{sec:rqs}

Here we outline the study's research questions.

\begin{enumerate}[label={\textbf{RQ\arabic*}},ref={RQ\arabic*}]
    \item\label{rq-rpro} \textbf{Can we replicate the results of \citet{original-study} on a similar dataset?}  The prior study analyzed over 2.7k Java projects. We want to know if it is possible to replicate 
    their results independently with our own analysis scripts and dataset (with as close to the same set of projects as possible).

    \item\label{rq-larger} \textbf{Do the observations and trends of \citet{original-study} still hold when analyzing a much larger set of Java projects?}  While the prior study looked at over 2.7k Java projects, do those trends still hold for a larger dataset with 89k Java projects (30x larger)?

    \item\label{rq-kotlin} \textbf{Do Kotlin programmers use method chaining in a way similar to Java programmers?}  To see if the trends observed for Java generalize to other languages, we first look at another JVM-based language: Kotlin.

    \item\label{rq-python} \textbf{Do Python programmers use method chaining in a way similar to Java or Kotlin programmers?}  To see if the trends observed for Java or Kotlin generalize to other, non-JVM based, languages, we next look at Python--a popular scripting language.

    \item\label{rq-extend} \textbf{Can we support the language extensions proposed by \citet{original-study}?}  Their study proposed extensions to the Java language to better encourage and support the use of method chains.  Are we able to support those recommendations with our larger and more diverse datasets?
\end{enumerate}

In the next section, we discuss our approach to investigate each of these research questions.

\section{Approach}
\label{sec:approach}

In this section we outline the approach used to answer our research questions.  First we discuss the data used.  Then we discuss how we query that data to find method chains.  Finally, we discuss the methodology used to analyze the query results.

\subsection{Datasets}

For each research question, we either built a new Boa~\cite{boa-website,boa} dataset or used one of the existing datasets.  In total we used 4 different Boa datasets.  All datasets were built from public GitHub repositories marked as non-forks.

All repositories were located and cloned during the summer of 2021, thus 2020 is the last full year of data analyzed.  Similar to the prior study, we use a starting year based on the dataset having at least 250 projects.  An overview of the datasets is shown in \tabref{tab:datasets}.

Duplicate files across projects, if they exist, are filtered out, retaining one of each duplicate set.  Leaving duplicates in could bias the results, as any file(s) that are highly duplicated would contribute their method chains multiple times.  Similar to \textcite{lopes17:_dejav}, we filter duplicates by collecting the hash of each file's AST, which ignores whitespace and comment differences between files, and keeping one copy of each unique hash.  The total number of duplicate files identified is shown in the last row of \tabref{tab:datasets}.  Note that for Java this is around 30\% of the files, but that result matches prior studies of duplication in Java datasets~\cite{lopes17:_dejav,allamanis19:_onward}.

\begin{table}[ht]
    \centering
    \caption{Overview of analyzed datasets}
    \label{tab:datasets}
\newcommand{\oldtabcolsepA}{\tabcolsep}
\renewcommand{\tabcolsep}{4pt}
\begin{tabular}{lrrrrr}
\toprule
{} &  Java Original &       Java &    Kotlin &    Python \\
\midrule
Projects                            &          2,659 &     89,088 &    26,273 &    98,202 \\
Files                               &      1,195,239 & 19,388,933 & 1,208,310 & 4,818,579 \\
\rowcolor{gray!20}\qquad Duplicates &         10,824 &  6,407,969 &    49,643 &   929,866 \\
\bottomrule
\end{tabular}
\renewcommand{\tabcolsep}{\oldtabcolsepA}
\end{table}

The \textbf{Java Original} dataset (``2021 Method Chains'' in Boa) was built by first collecting the list of all projects from the original study's data~\cite{original-data}.  We then attempted to clone each from GitHub.  During that process we determined some were forks, which we excluded.  Renamed projects were cloned using their new names.  There were also 19 projects not available on GitHub.  We attempted to locate those on the Software Heritage Archive~\cite{sha} and found 13.  This gave us 2,659/2,756 (96.48\%) of the projects from the original study.

The original study chose the projects based on being in the daily top-1000 most starred projects in a small time window in 2019.  Thus, the projects actually have star counts ranging from 1 to over 100k.  Since we want a direct comparison to the prior paper, we do not filter this dataset.

The \textbf{Java} dataset (``2022 Jan/Java'' in Boa) contains projects indicating Java as primary language (highest percentage of code is Java) on GitHub and sorting based on star counts before cloning.  Cloning stopped when we ran out of space.  Note that while we did not set any threshold, since the cloning was sorted based on star counts, all Java projects have at least 10 stars.

The \textbf{Kotlin} dataset (``2021 Aug/Kotlin'') contains projects indicating Kotlin as primary language.  At the time of crawling, this represented almost every Kotlin project on GitHub.  Since the other three datasets were using a notion of popularity, we filtered this dataset based on star counts and kept projects with at least 2 stars (to avoid projects with only self-stars).  We considered using a higher cutoff such as 10 or 8 (to mirror the Java or Python datasets), but decided against it as filtering with these higher thresholds leads to a substantially smaller dataset.  We opted to keep the dataset size the same magnitude, to avoid making the datasets imbalanced.

The \textbf{Python} dataset (``2022 Feb/Python'') was built with projects indicating Python as primary language and then sorting based on star counts.  Again, note that while we did not set any threshold, since the cloning was sorted based on star counts, all Python projects have at least 8 stars.

\subsection{Finding Method Chains}

To mine method chains from the datasets, we needed to write several Boa queries.  \figref{fig:chainlen} shows a helper function \lstinline|chainlen| that, when given a method call expression, returns the length of that method chain.

\begin{figure}[ht]
\begin{Boa}
chainlenhelper := function(e: Expression) : int {
  if (len(e.expressions) > 0 &&
      e.expressions[0].kind == ExpressionKind.METHODCALL)
    return 1 + chainlenhelper(e.expressions[0]);
  return 0;
};

chainlen := function(e: Expression) : int {
  return 1 + chainlenhelper(e);
};
\end{Boa}
  \vspace{-0.5em}
  \caption{Boa functions to measure method chain lengths}
  \label{fig:chainlen}
\end{figure}

Method chains in Boa are represented in the AST nested, meaning the tree root is the last call in the chain and as you traverse down the tree you find the earlier call(s) in the chain.

\begin{figure}[ht]
\begin{Boa}
before e: Expression -> {
  curlen := 0;
  lens: stack of int;
  visit(e, visitor {
    before e: Expression -> {
      if (e.kind != ExpressionKind.METHODCALL) {
        curlen = 0;
      } else {
        if (e.method != "super" &&
            e.method != "<init>" &&
            curlen == 0) {
          curlen = chainlen(e); # found new chain
        }

        push(lens, curlen);
        foreach (i: int; e.method_args[i]) {
          curlen = 0;
          visit(e.method_args[i]);
        }
        curlen = pop(lens);
        if (curlen > 0) curlen -= 1;

        foreach (i: int; e.expressions[i])
          visit(e.expressions[i]);
        stop;
      }
    }
  });
  stop;
}
\end{Boa}
  \vspace{-0.5em}
  \caption{Boa query snippet to locate method chains}
  \label{fig:findchains}
\end{figure}

\figref{fig:findchains} shows the main query.  For every expression found, a sub-visit is needed to locate the chain(s).  This is because there could be a path in the AST with several method calls, but with other expressions following the dot operator, such as a field access.  We also don't want to locate a chain and then move one call down the AST and accidentally report another (sub)chain.  This is handled with the \lstinline|curlen| counter, where 0 means we found a new chain.

The accuracy of the method chain locating queries was verified using manually developed test cases.  This was especially important to verify in Python, where chains that span multiple lines require either backslashes or the entire chain must be enclosed in parentheses.\footnote{\url{https://stackoverflow.com/questions/48863091/pep8-chained-methods}}

\subsection{Analysis Approach}

We analyze the data with several Python scripts and utilize the Pandas library.  We first convert the text output from Boa into a CSV format, then import into Pandas for further processing.  File deduplication occurs after loading from CSV.

The scripts process each dataset and generate $f_n$ and $r$ values using equations \ref{eq:fn} and \ref{eq:r}.  Similar to the prior study, we then scatter plot the $f_n$ values for the first/last years for each dataset to visualize how the frequency of method chaining for different chain lengths has changed over time.  We also show a bar plot of the overall $r$ values over that time range, to see if there is an increasing or decreasing trend.

Similar to the original study, we then investigate the distribution of varying lengths of methods chains, categorizing each chain into ``short'' (less than or equal to $n$, where $U_n$ is closest to 50\%), ``long'' (where $U_n$ is closest to 5\%), and ``extra long'' (where $U_n$ is less than 5\%).  The categories for the oldest year are used to compare the distribution of chain lengths across all years for each dataset.

Finally, similar to the original study, we look at testing vs non-testing code as our experience with Java code tells us testing code often looks different.  We suspect testing behavior across languages may vary as well.  
Observing any differences also helps to understand if method chaining supports the specific goals of testing code.  For this, we mark each file as a test based on the lowercase path containing a sub-string ``test'' or if the file imports one of the top testing libraries/modules.  Here we present the results as both scatter plots of their $f_n$ values and bar plots of their $r$ values over time, so we can observe if the trends changed.

\section{Results}
\label{sec:results}

In this section we detail the results of our empirical study.

\subsection{\ref*{rq-rpro}: Can we replicate prior results on a 
similar dataset?}

First, we wanted to verify the analysis scripts we created to generate tables/charts (similar to the original paper) worked as expected.  To verify them, we used the original data files~\cite{original-data} provided by the prior study~\cite{original-study} and processing scripts they provided directly to us.  We were able to successfully use their scripts and data to reproduce the results in their paper.

\findings{
    We were able to reproduce the prior paper's results using their data and scripts.
}

We then converted their \texttt{data.txt} results file by loading it into a Pandas \texttt{DataFrame} that our analysis scripts operate on.  The results showed that our analysis scripts were able to correctly reproduce the graphs from their Figures 3, 4, 5, and 6 and their Table 1.  Thus we feel confident our analysis scripts were written correctly.

During this process we did identify a single anomalous result in the test vs non-test scatter plot.  After investigating, we determined that while we searched for ``test'' in the file path in a case-insensitive manner, they appeared to look only for lowercase matches and thus missed a single data point
.  However, when they calculated the percentage of extra-long chains in testing code out of all extra-long chains, they performed a case-insensitive search.

Next, we aimed to replicate the prior study by using a new (but similar) dataset we built, which we call Java Original.  This dataset was built using the same set of projects from the original study, but cloned at a different point in time (thus they are not 100\% identical).  We then used Boa queries to mine our dataset and identify method chains.  We hoped the results would be very close to the ones observed in the prior steps, indicating the Boa queries correctly mined the method chains.

However we noticed quite a few differences between these results and the chains provided by the previous study.  After quite a bit of manual analysis, we were able to determine the Boa queries were identifying a lot of chains that were not included in the original paper's data file.  We were able to identify a few common patterns among the missed data and then confirmed with the original authors that there were some bugs in their script used to mine the method chains (that script is not provided in the replication package).

We also identified some inconsistencies in the reported number of projects, as while 2,814 projects were reported, only 2,756 had Java source files in the date range studied.  Additionally, 21 of those projects were unintentionally kept despite being forks of other projects in the dataset (and thus, exact duplicates).  We also identified 10,824 files that were duplicated across (non-forked) projects that were unintentionally included in their data.  Finally, it seems like the method used to snapshot projects by year may have had some issue, as we were able to identify a file with method chains that was deleted in 2015, modified in 2014, but the mined data contains no chains past 2011 for some reason.

We were able to communicate with the lead author and confirm most of the problems identified~\cite{original-errata}.  Interestingly, since some of the errors resulted in fewer chains found and others resulted in more chains (from dupes), the total number of chains found was almost the same: 152,161,181 from them vs 152,783,246 from us.  The analysis on the data also resulted in the same trends with very minor variations (we do not show them here - but full results are in our replication package).

\findings {
    Despite identifying several irregularities, we independently replicated the prior study's main results, showing that method chaining is (increasingly) popular in Java code.
}

We conclude by answering the research question: we were able to successfully replicate the prior study's results.

\begin{figure}[t]
    \centering
    \includegraphics[width=0.49\linewidth]{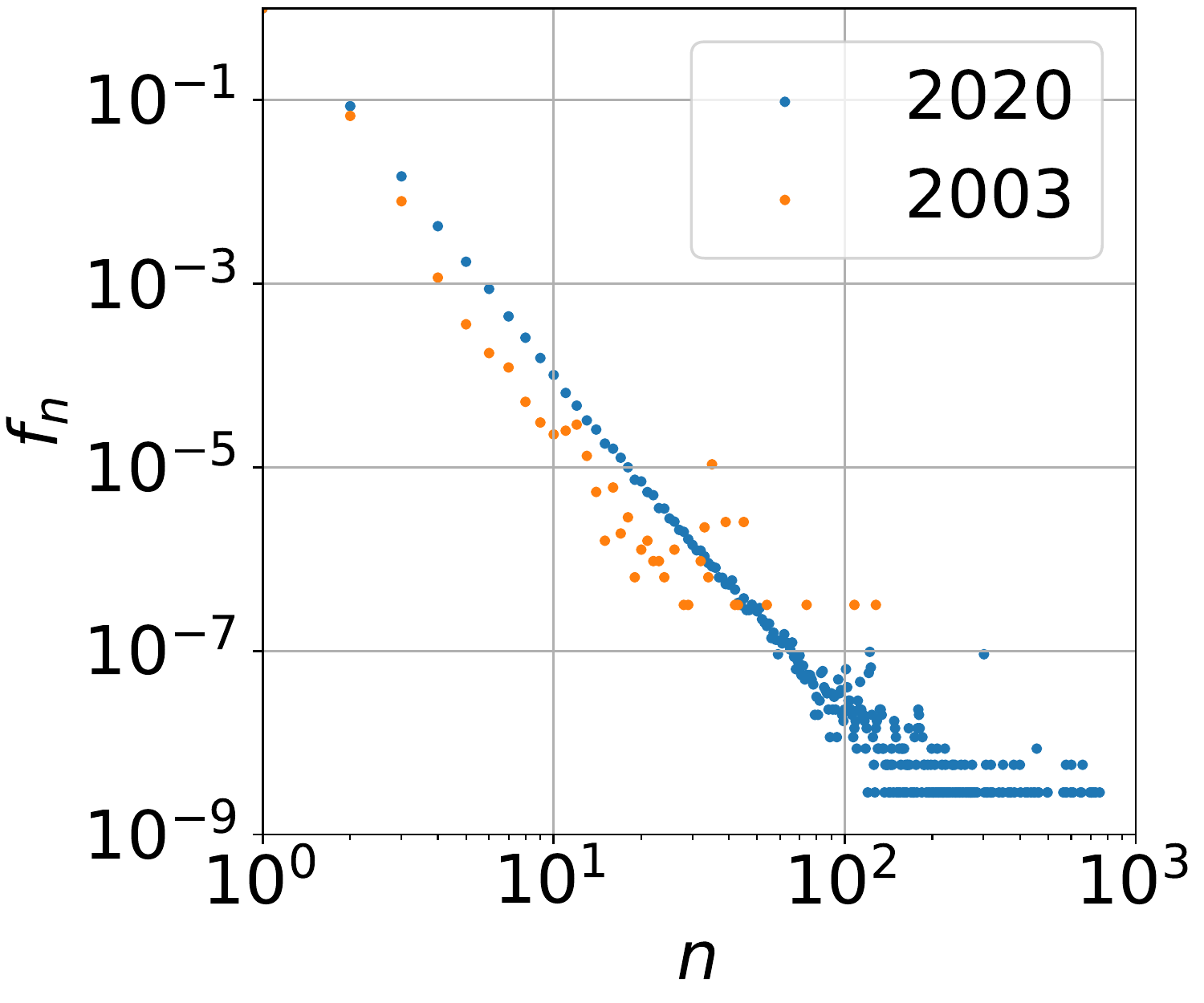}
    \includegraphics[width=0.49\linewidth]{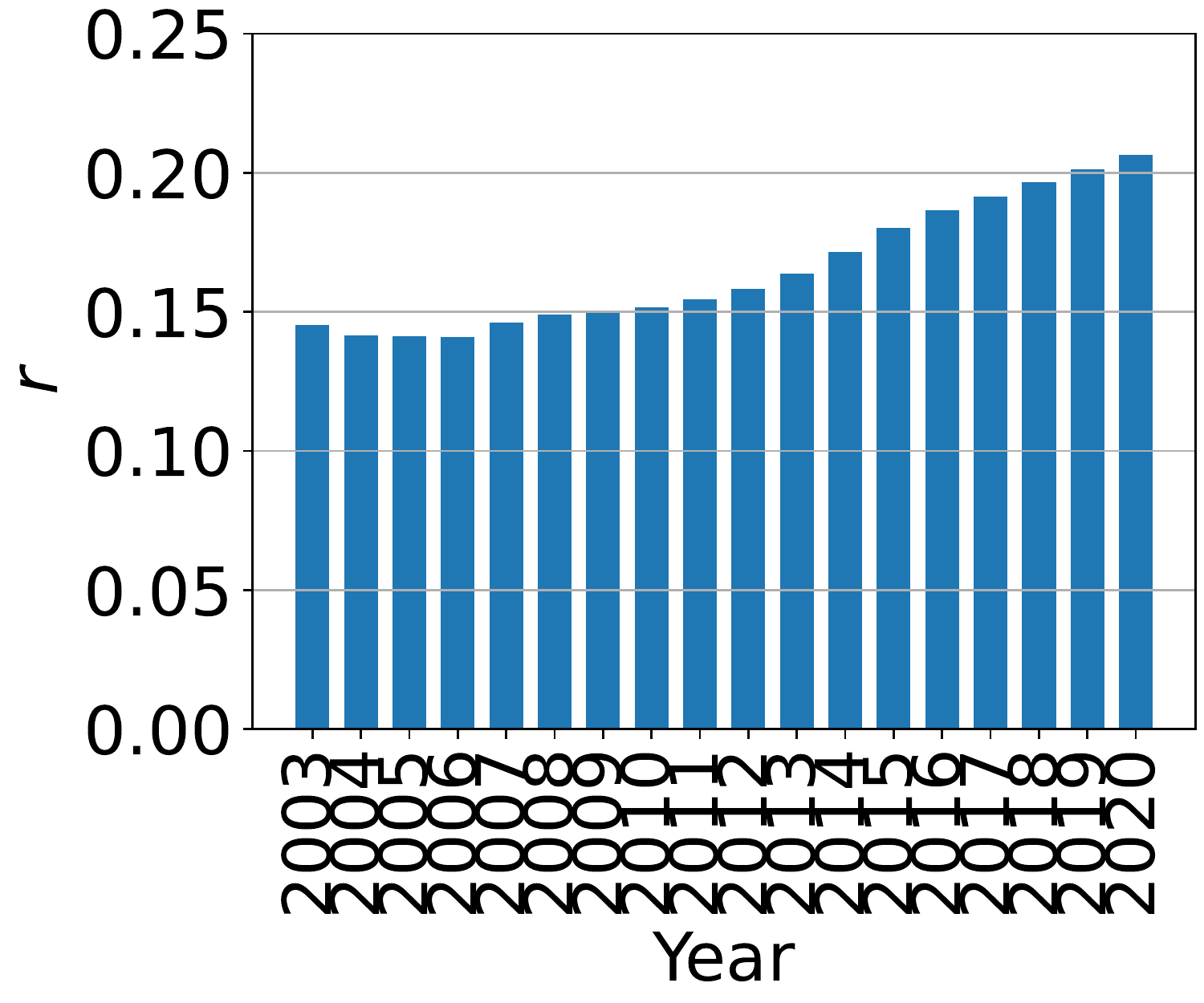}
    \caption{$f_n$ and $r$ values (Java)}
    \label{fig:java:scatter}
\end{figure}

\begin{figure}[t]
    \centering
    \includegraphics[width=0.49\linewidth]{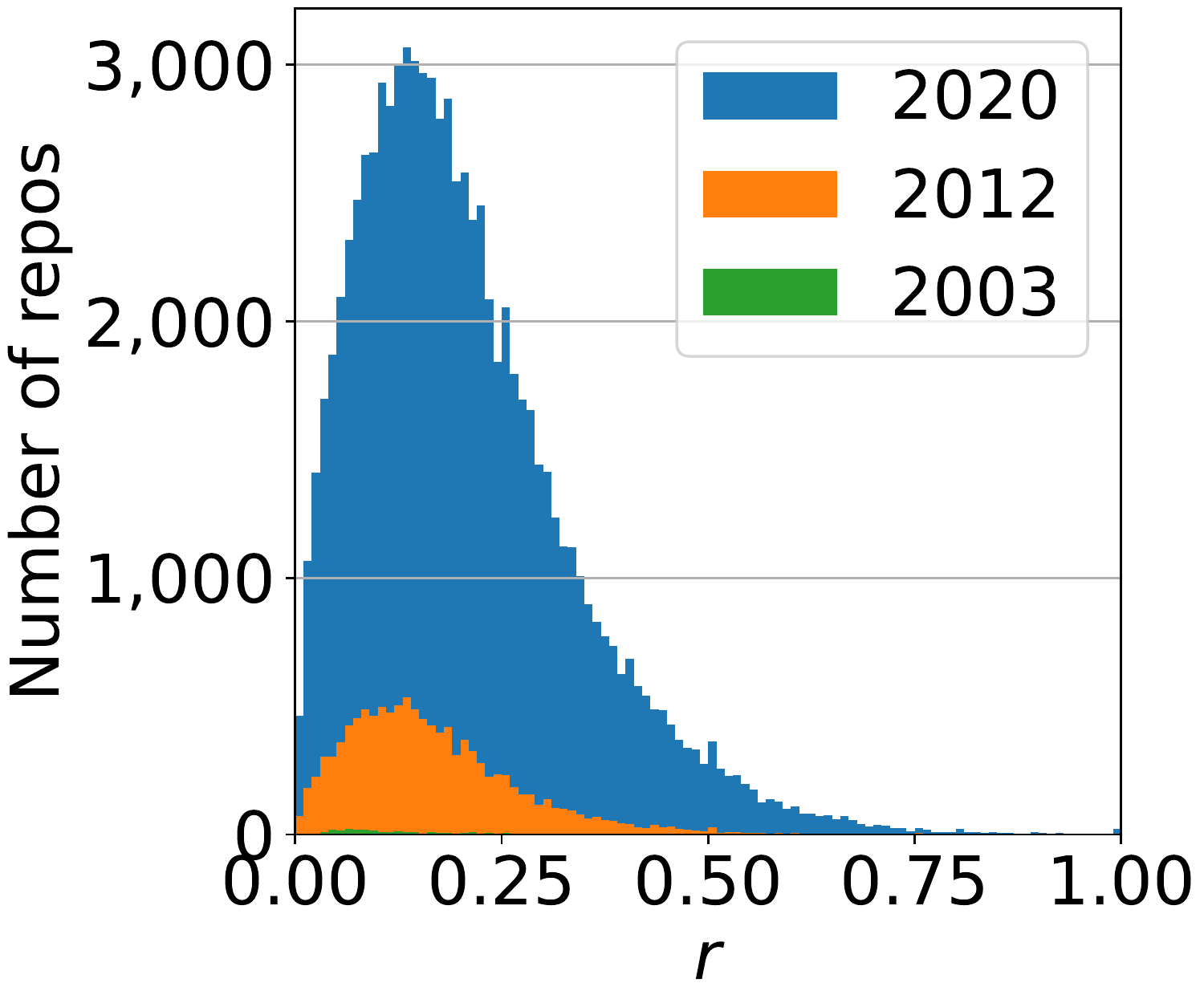}
    \includegraphics[width=0.49\linewidth]{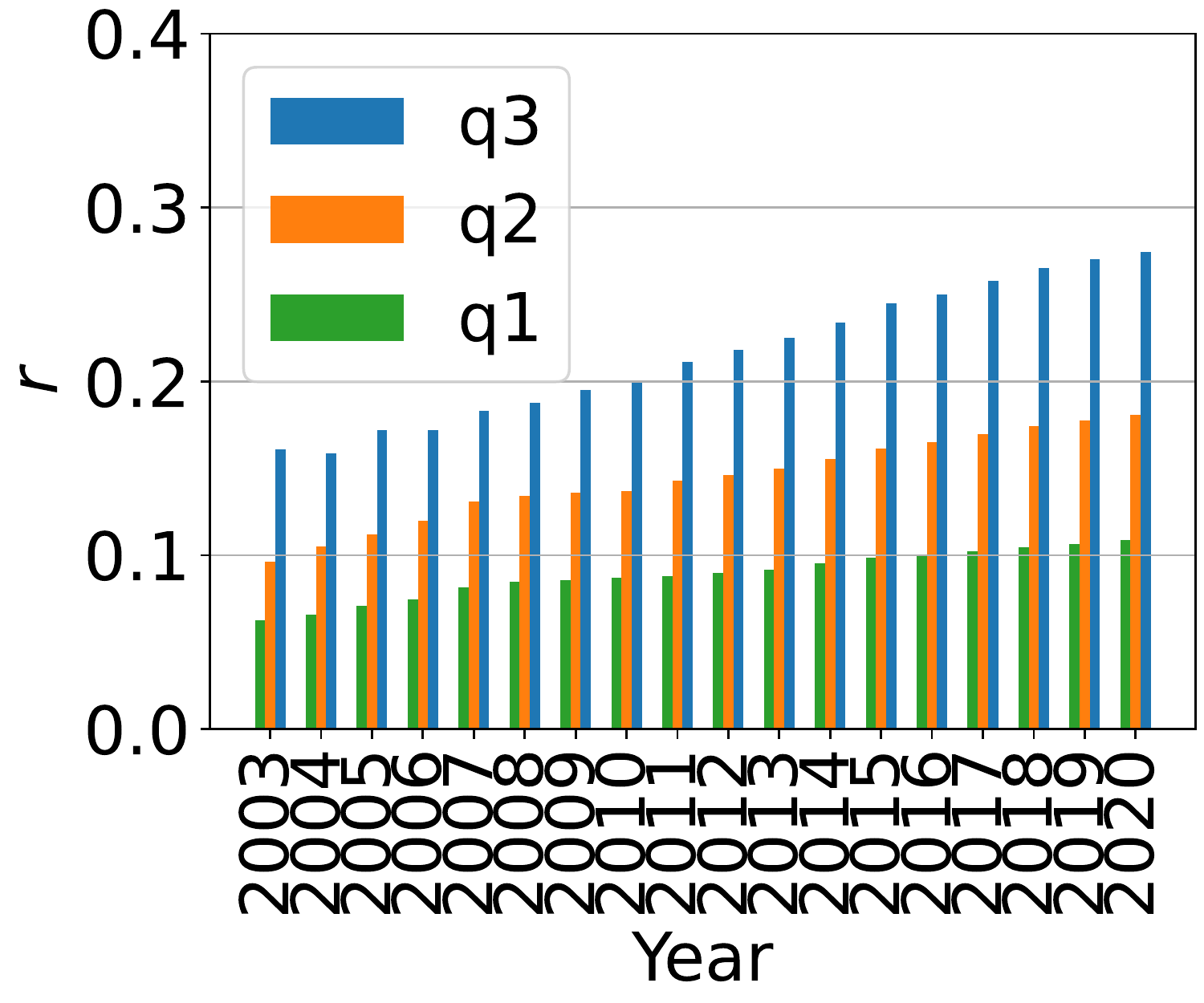}
    \caption{Distribution and trend of $r$ values per project (Java)}
    \label{fig:java:hist}
\end{figure}

\subsection{\ref*{rq-larger}: Do the prior study's observations and trends still hold when analyzing a much larger set of Java projects?}
\label{subsec:javafull}

Since we were able to replicate the prior study's results, we wanted to first see if those results generalize to a larger set of Java projects.  \figref{fig:java:scatter} shows the $f_n$ and $r$ values over the total dataset for the years 2003--2020.  Note the scatter plot uses log-scale for both axes.  Similar to the results for the Java Original dataset (not shown), these results indicate increasing use of method chains in Java.  This observation is further confirmed when viewing the histogram and quartiles plot in \figref{fig:java:hist}.  From 2003--2020, all three quartiles increase by 3--10\%.

\findings {
    The increased use of method chains observed in the original, smaller study are also observed in a dataset 35x larger.
}

\begin{figure}[t]
    \centering
    \includegraphics[width=0.49\linewidth]{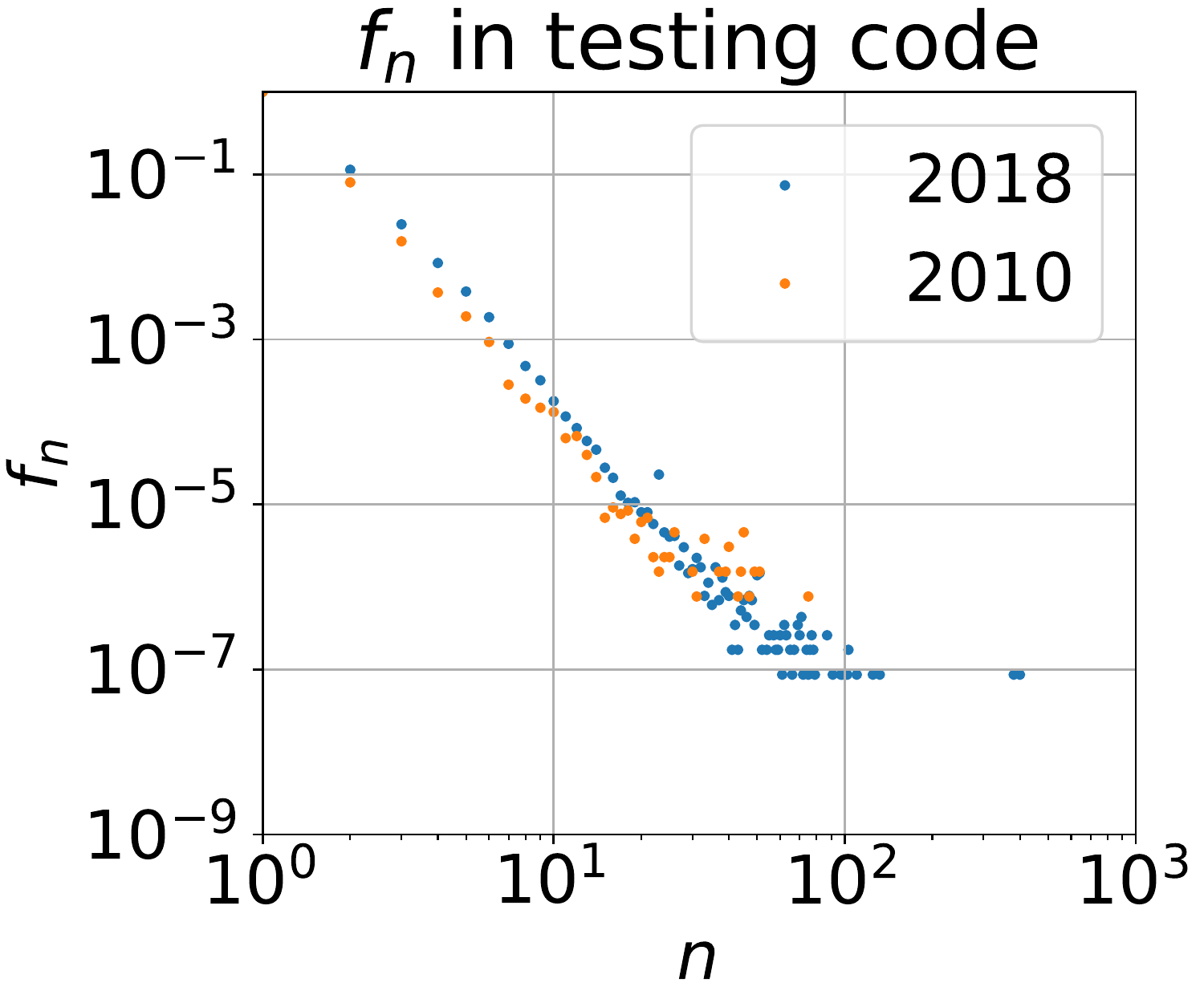}
    \includegraphics[width=0.49\linewidth]{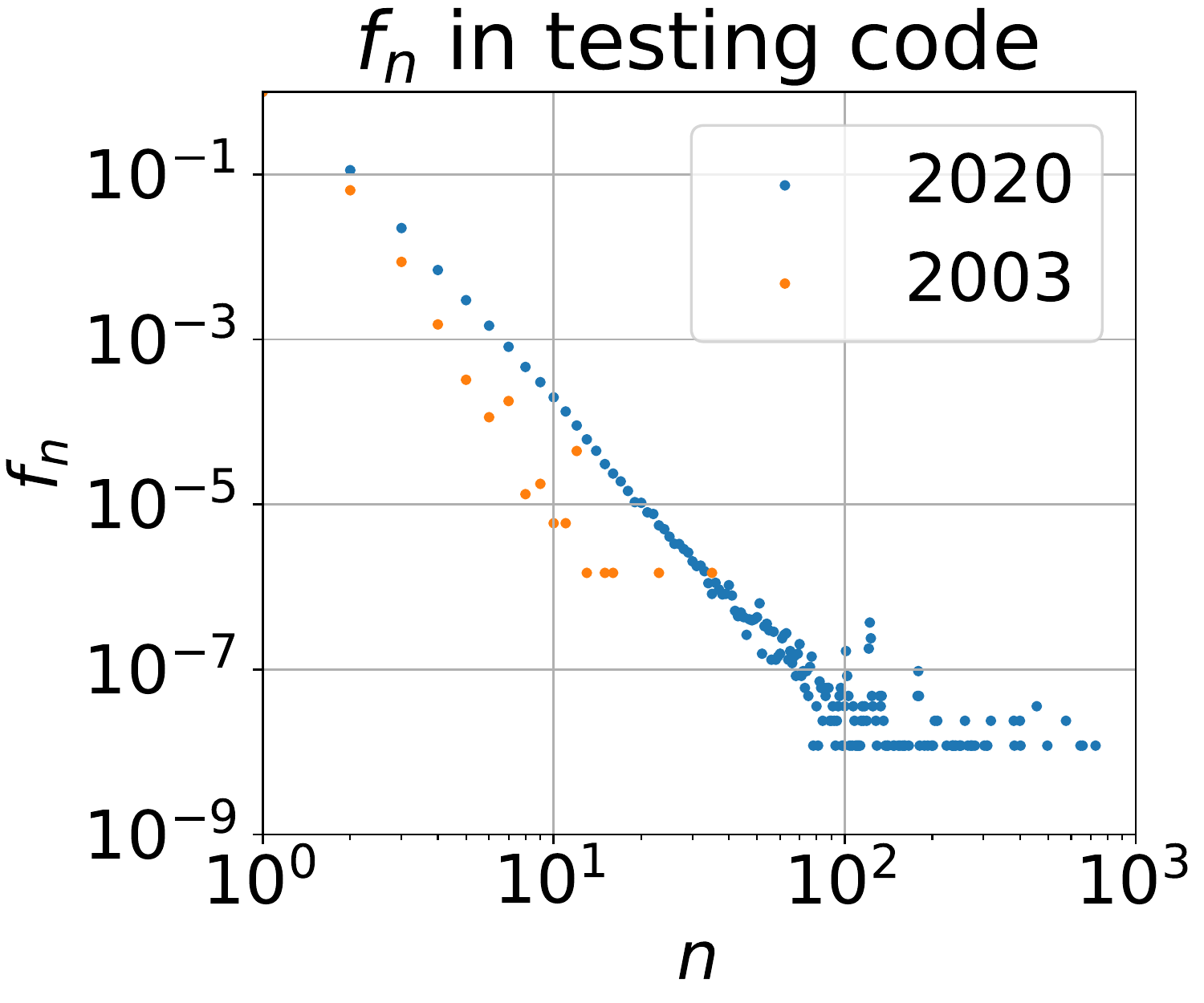}
    \vspace{-1.5em}
    \caption{$f_n$ of testing code (Java Original left, Java right)}
    \label{fig:java:comparetest}
\end{figure}

\begin{figure}[t]
    \centering
    \includegraphics[width=0.49\linewidth]{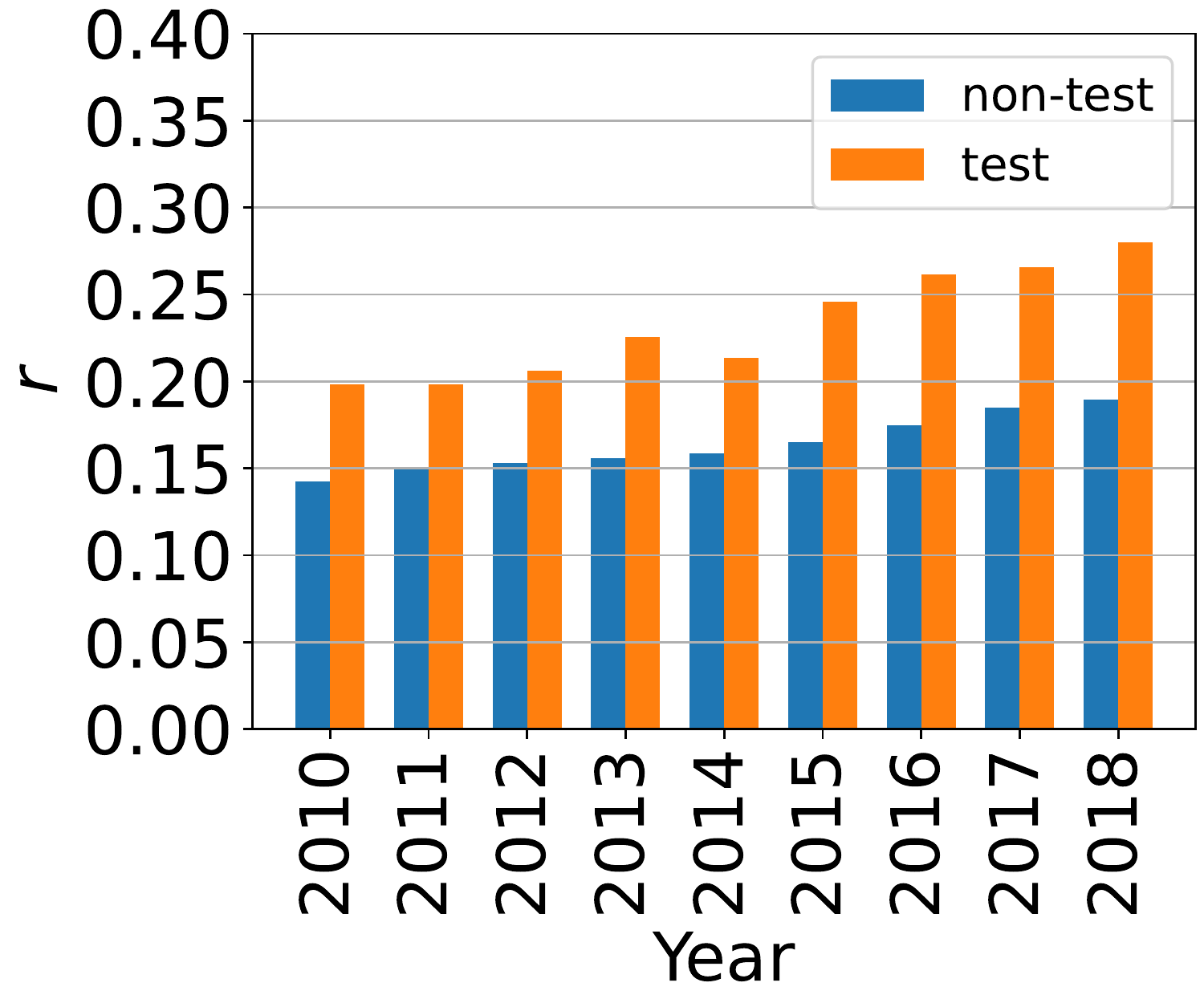}
    \includegraphics[width=0.49\linewidth]{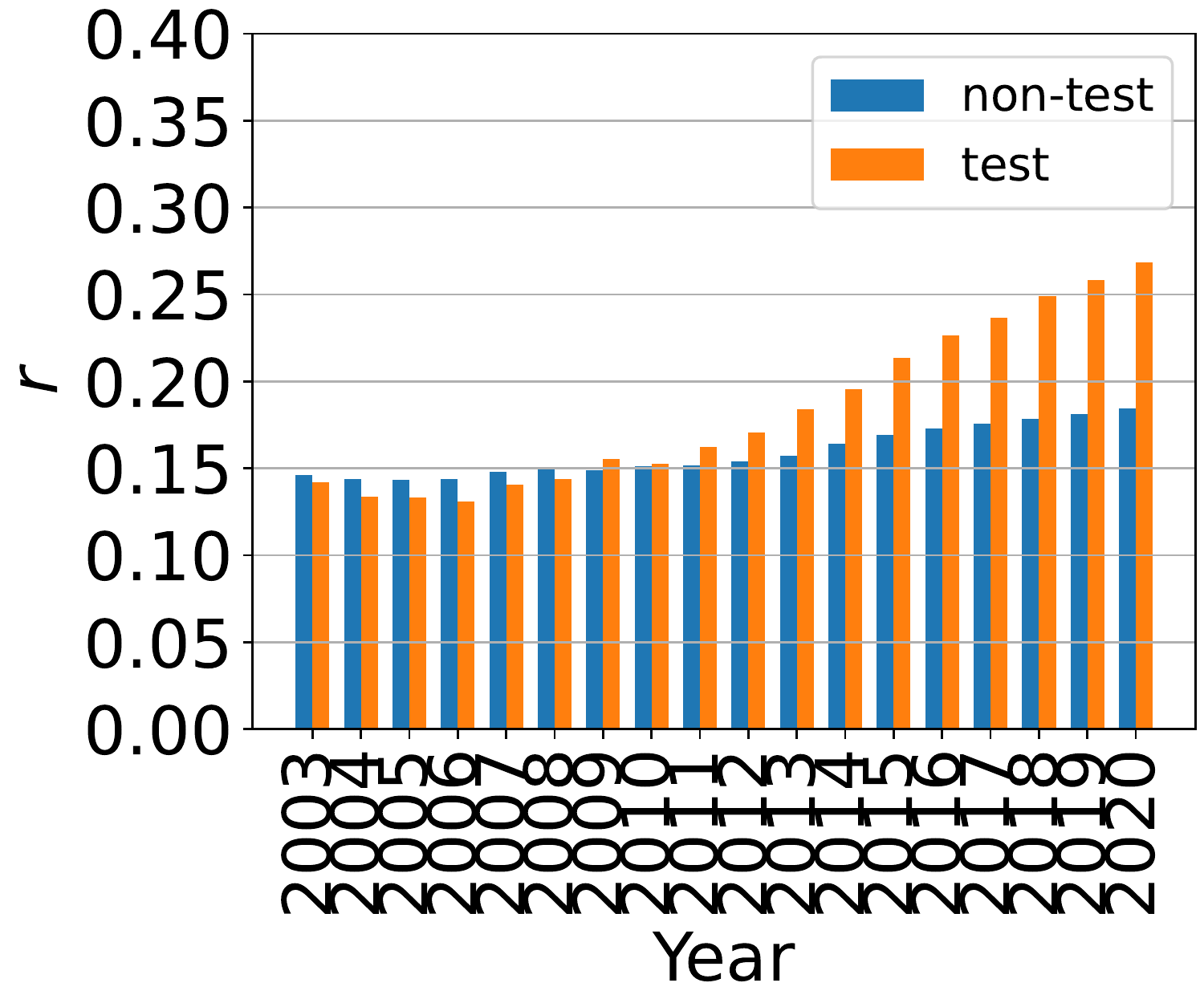}
    \vspace{-1.5em}
    \caption{Comparing $r$ values for non-testing vs. testing code (Java Original left, Java right)}
    \label{fig:java:compare}
\end{figure}

We do however observe some minor differences from the Java Original dataset.  When looking at method chaining in testing code, \figref{fig:java:comparetest} shows the scatter plots of the $f_n$ values with the Java dataset on the right.  Despite of the fact the years are different (first/last year of each dataset), when we looked at the non-testing plots they were quite similar but the testing code seems to show different behavior for the year 2003.  This was confirmed looking at the $r$ plots shown in \figref{fig:java:compare}.

It appears that in older Java projects, there was more method chain use happening in non-testing than in testing code.  This was not observed in the original study, possibly because their data started in 2010 when the trends flipped.  This seems to indicate the growth in the use of method chains in testing code is more pronounced than previously thought.

\findings{
    Method chaining in Java originally occurred more in non-testing code before 2011. Since then, method chaining occurs more in testing code and seems to be growing faster.
}

\begin{table}[ht]
    \centering
    \caption{Groups of method chains (Java)}
    \label{tab:java:chaingroups}
\newcommand{\oldtabcolsepA}{\tabcolsep}
\renewcommand{\tabcolsep}{2pt}
\begin{tabular}{lrrrr}
\toprule
{} &                 Short &                Long &          ExtLong \\
\midrule
Length range      &      1 < len. $\le$ 4 &       4 < len. < 23 &    23 $\le$ len. \\
\# chains in 2020 &  36,557,801 (96.46\%) &  1,329,866 (3.51\%) &  12,907 (0.03\%) \\
\# chains in 2014 &  12,207,859 (97.99\%) &    247,078 (1.98\%) &   2,873 (0.02\%) \\
\# chains in 2005 &     461,495 (99.13\%) &      3,974 (0.85\%) &      89 (0.02\%) \\
\# chains in 2003 &     240,932 (98.86\%) &      2,695 (1.11\%) &      79 (0.03\%) \\
\bottomrule
\end{tabular}
\renewcommand{\tabcolsep}{\oldtabcolsepA}
\end{table}

In \tabref{tab:java:chaingroups} we group method chain uses into categories based on lengths, from short, to long, to extra long.  Here we also observe differences with the prior results.  Compared to the prior study, our larger dataset has a higher percentage of long and extra long chains.

\begin{figure}[t]
    \centering
    \raisebox{-.5\height}{\includegraphics[width=0.5\linewidth]{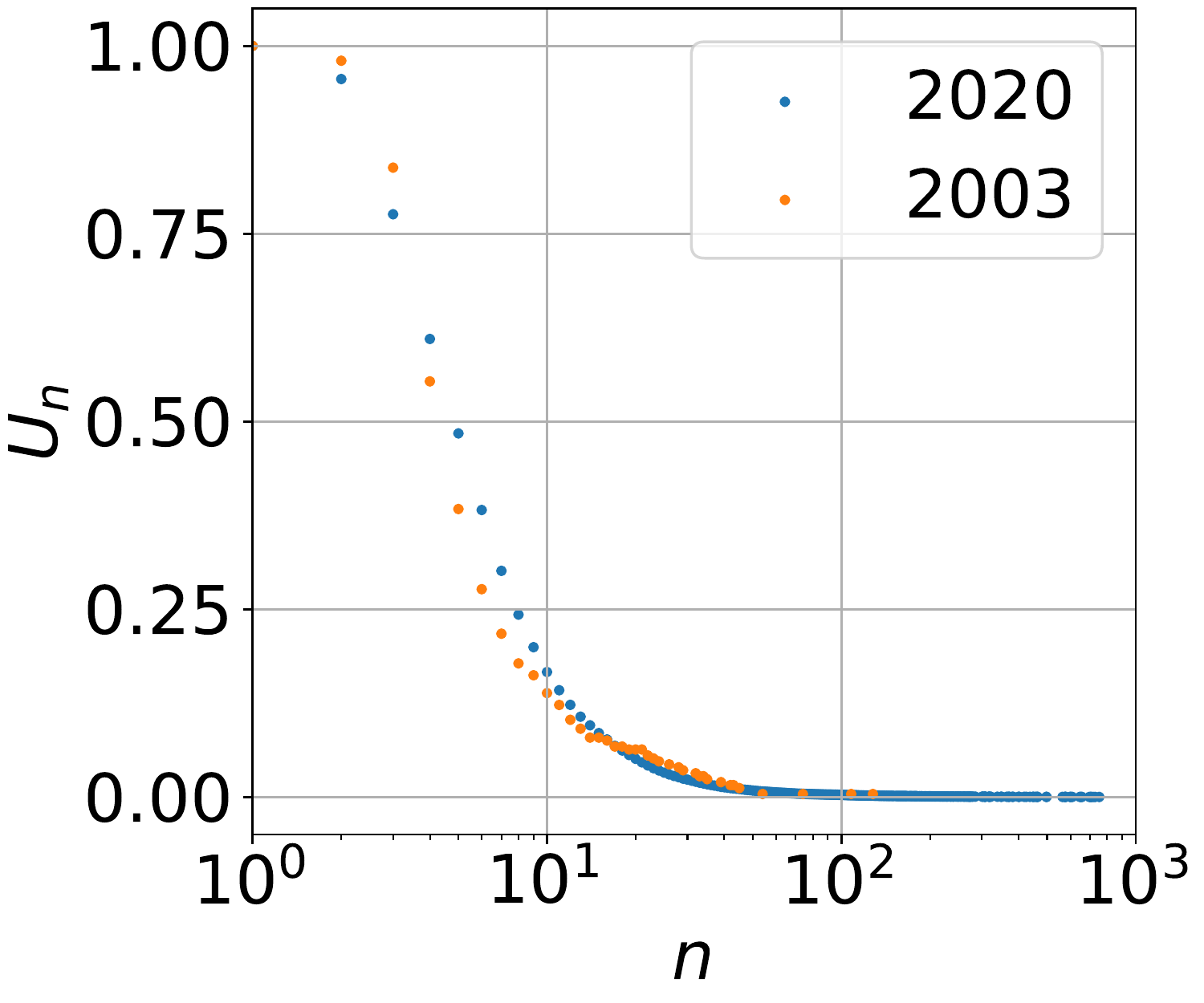}}
\newcommand{\oldtabcolsep}{\tabcolsep}
\renewcommand{\tabcolsep}{2pt}
\begin{tabular}{lrrrr}
\toprule
{} & $U_n$ in 2020 & $U_n$ in 2003 \\
$n$ &               &               \\
\midrule
1   &      100.00\% &      100.00\% \\
8   &       24.28\% &       17.79\% \\
9   &       19.94\% &       16.21\% \\
41  &        1.24\% &        1.58\% \\
42  &        1.18\% &        1.58\% \\
\bottomrule
\end{tabular}
\renewcommand{\tabcolsep}{\oldtabcolsep}
    \caption{Ratio of projects containing chains longer than or equal to $n$ (Java)}
    \label{fig:java:ratios}
\end{figure}

When observing the ratios shown in \figref{fig:java:ratios}, where $U_n$ is the ratio of repositories containing at least one chain of length $n$, we observed some very large differences.  For most values of $n$, the ratios shown here are about 2--4x smaller than the ratios from the prior study.  We do however observe increasing ratios from 2003--2020, similar to the prior study.

\findings{
    Similar to the prior study, we found that method chains in Java are mostly short and adopted more over time.
}

Finally, we looked at what some of the more popular libraries are that produced the extra long chains.  First, we wrote a Boa query that attempts to infer the static type of every identifier in the code as well as the static type returned from method calls.  We then ran a query to find what type the first method call in each extra long chain was and then grouped them by library.  The results are shown in \tabref{tab:java:libraries}.

\begin{table}[ht]
    \centering
    \caption{Popular libraries producing extra long chains (Java)}
    \label{tab:java:libraries}
    \newcommand{\oldtabcolsep}{\tabcolsep}
    \renewcommand{\tabcolsep}{4pt}
\begin{tabular}{lrrrr}
\toprule
{} &  Count & Percentage \\
Package                                              &        &            \\
\midrule
java.lang.StringBuilder                              &  2,366 &    16.24\% \\
org.springframework.security...builders.HttpSecurity &    816 &     5.60\% \\
com.google.common.collect.ImmutableMap               &    587 &     4.03\% \\
java.lang.StringBuffer                               &    519 &     3.56\% \\
javax.swing.GroupLayout                              &    445 &     3.05\% \\
vm.runtime.defmeth.shared.builder.TestBuilder        &    414 &     2.84\% \\
org.elasticsearch.common.xcontent.XContentFactory    &    214 &     1.47\% \\
com.google.common.base.MoreObjects                   &    211 &     1.45\% \\
org.assertj.db.api.Assertions.assertThat             &    202 &     1.39\% \\
org.apache.avro.SchemaBuilder                        &    153 &     1.05\% \\
\bottomrule
\end{tabular}
    \renewcommand{\tabcolsep}{\oldtabcolsep}
\end{table}

The prior study had a small sample to deal with (only 71 extra long chains) while we analyzed almost 13k.  Given that size difference, it is good to see all three of their most commonly identified libraries are in our results as well.  In addition to Guava and the Java standard library, we also identified the Spring framework in our top 5 list.

\findings{
    Both our study and the prior study found quite a few libraries account for a large number of the extra long method chains observed.  In our case, 1/3 of the extra long chains came from just five libraries.
}

We conclude by answering the research question: the prior study's observations and trends still hold on a larger dataset.

\subsection{\ref*{rq-kotlin}: Do Kotlin programmers use method chaining in a way similar to Java programmers?}
\label{subsec:kotlin}

\begin{figure}[t]
    \centering
    \includegraphics[width=0.49\linewidth]{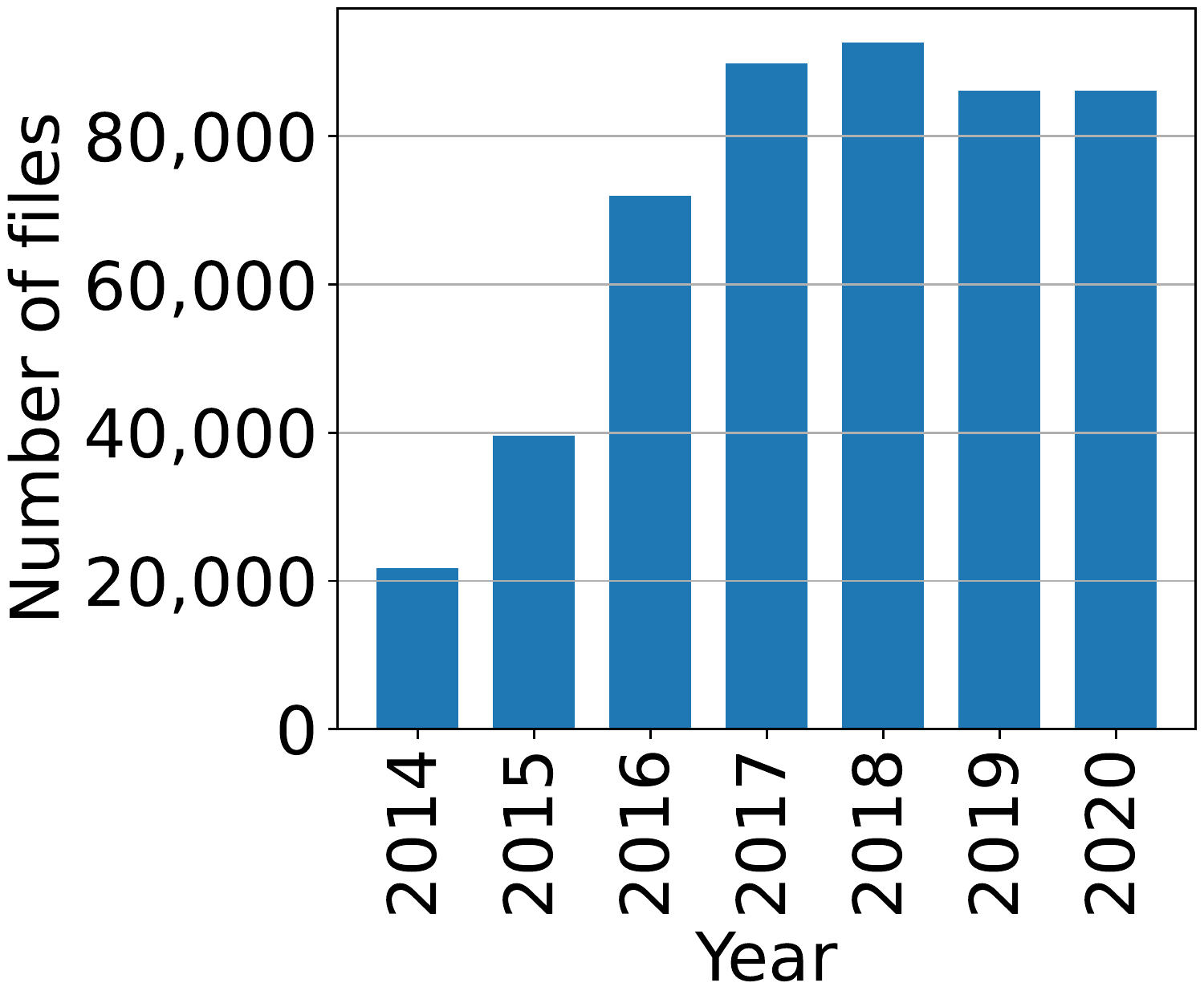}
    \includegraphics[width=0.49\linewidth]{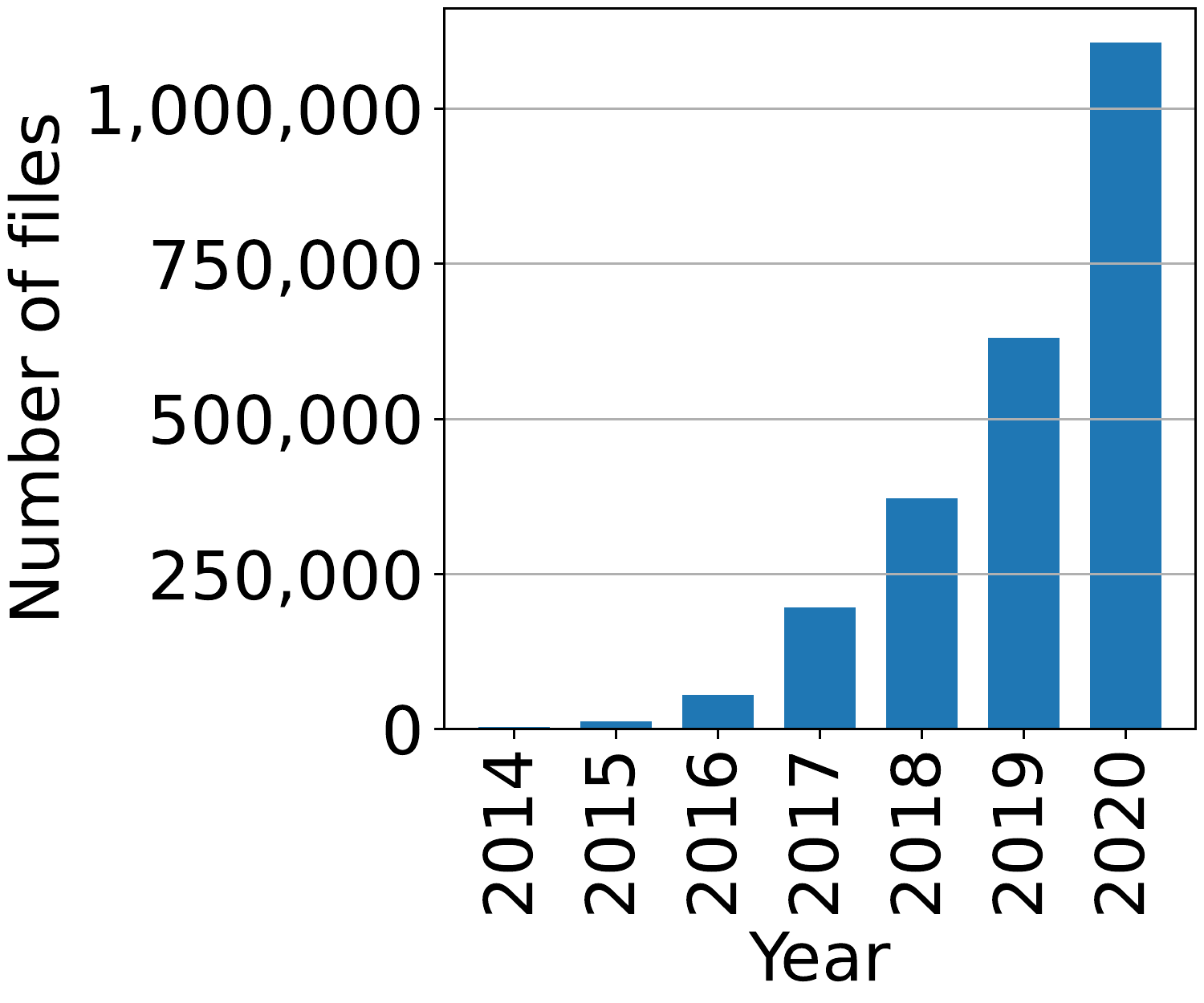}
    \caption{Kotlin project file counts (Java left, Kotlin right)}
    \label{fig:kt:filecounts}
\end{figure}

So far we have only investigated how method chaining was used in Java projects.  Our next two research questions try to see if the trends are similar for other programming languages.  First, we look at a JVM-based language called Kotlin that is Android's preferred programming language.  Kotlin was designed to interoperate with Java, and many Kotlin projects actually contain both Kotlin and Java source files.  It is thus important to know the distribution of those files over time, which we show in \figref{fig:kt:filecounts}.

As can be seen, the number of Kotlin files increases rapidly across the years, while the number of Java files starts increasing and then around 2018 starts decreasing.  We suspect this is due to people becoming more comfortable with the Kotlin language and writing less and less code in Java over time, as 82\% of the projects are Android projects (where the default language used to be Java and is now Kotlin). It is also worth mentioning that 37 projects switched their default language from Java to Kotlin.

\begin{figure}[t]
    \centering
    \includegraphics[width=0.49\linewidth]{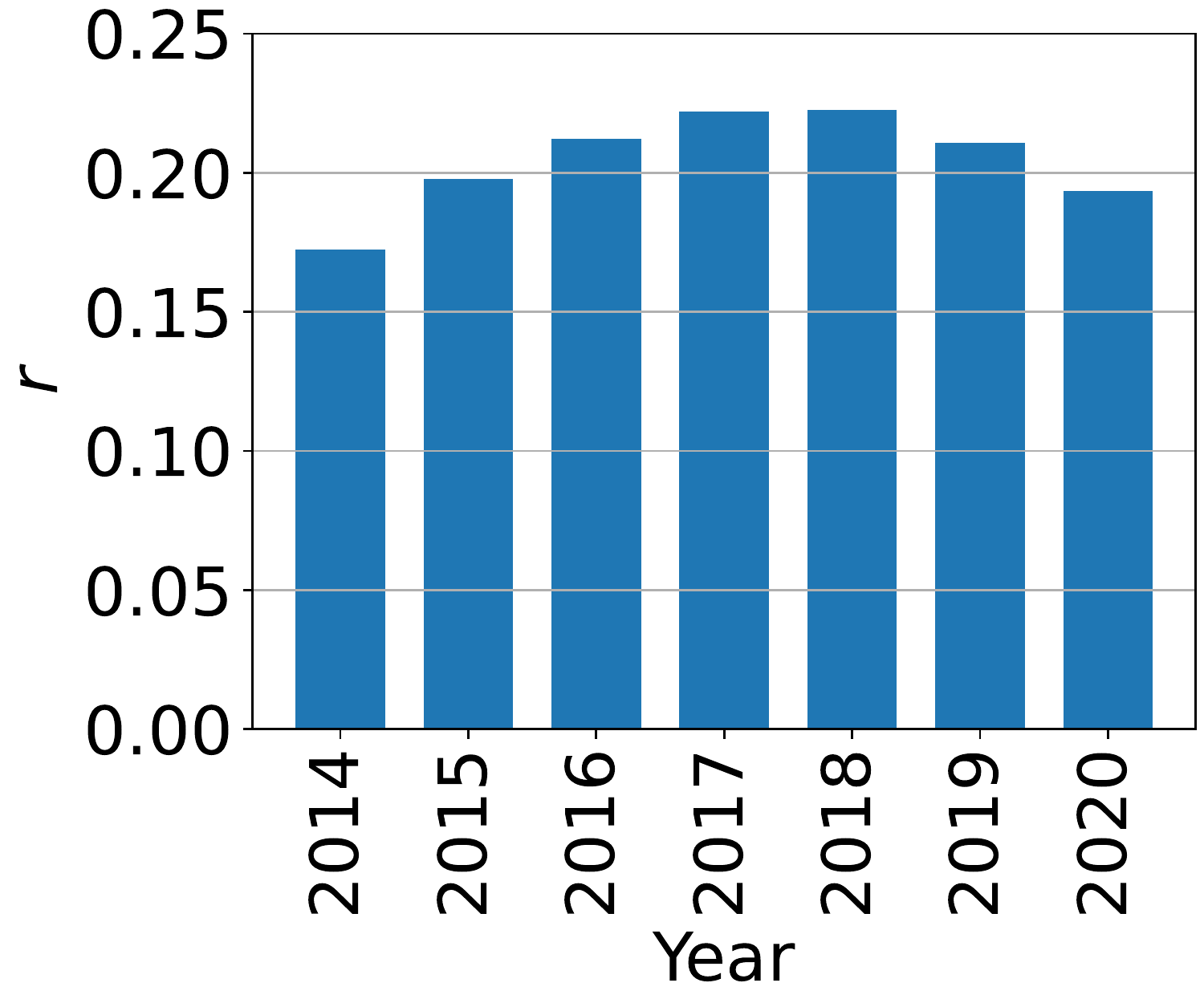}
    \includegraphics[width=0.49\linewidth]{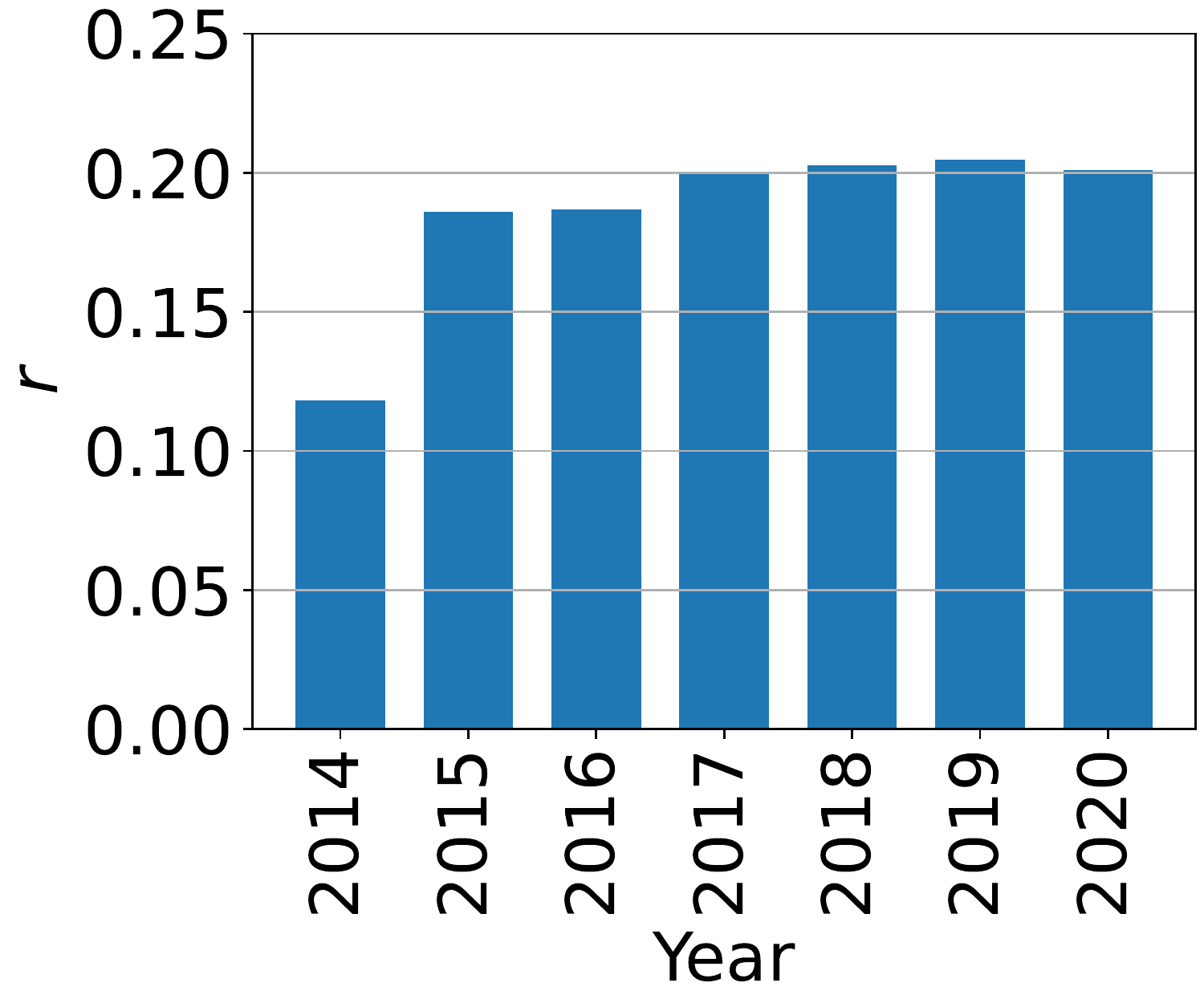}
    \caption{Kotlin project $r$ values (Java left, Kotlin right)}
    \label{fig:kt:compare-r}
\end{figure}

\figref{fig:kt:compare-r} shows the $r$ values for the Kotlin dataset, broken down by Java files on the left and Kotlin files on the right.  Both actually show similar trends in that they start increasing, then decrease the later years.  Both are also similar in terms of magnitude, meaning at least within a Kotlin project people seem to utilize method chaining about equally in both languages.  This is a bit surprising, given some of the additional language features Kotlin has that might avoid the need to chain.  We investigate some language support later in \secref{subsec:extend}.

\findings{
    The use of method chaining in Kotlin is \textbf{not} increasing, unlike Java.  Within a Kotlin project, developers seem to use method chaining similarly across the two languages.
}

When observing how the $r$ value quartiles changed in Kotlin projects, shown in \tabref{tab:kt:compare-quartiles}, we can see differing trends.  First, for Java source files we observe the first quartile going down and the third going up.  This means that in 2020 there is more variance (wider spread of quartiles) of method chain use within Java files.  We do not observe that for Kotlin files, where both the first and third quartiles show an overall increasing trend.

\begin{table}[ht]
    \centering
    \caption{Kotlin project $r$ quartile changes 2015--2020}
    \label{tab:kt:compare-quartiles}
\begin{tabular}{rrr}
\multicolumn{3}{c}{\textbf{Java Files}} \\
\toprule
1st & 2nd & 3rd \\
Quartile & Quartile & Quartile \\
\midrule
     -0.87\% &      +0.28\% &      +7.39\% \\
\bottomrule
\end{tabular}
\begin{tabular}{rrr}
\multicolumn{3}{c}{\textbf{Kotlin Files}} \\
\toprule
1st & 2nd & 3rd \\
Quartile & Quartile & Quartile \\
\midrule
     +3.69\% &      +5.11\% &      +5.38\% \\
\bottomrule
\end{tabular}
\end{table}

\begin{table}[ht]
    \centering
    \caption{Ratio of source files in Kotlin projects containing chains longer than or equal to $n$}
    \label{fig:kt:compare-ratios}
\begin{tabular}{rrrp{0.1cm}rr}
    & \multicolumn{2}{c}{\textbf{Java Files}} & & \multicolumn{2}{c}{\textbf{Kotlin Files}} \\
\toprule
{}  & $U_n$ in 2020 & $U_n$ in 2014 & &  $U_n$ in 2020 & $U_n$ in 2014 \\
$n$ &               &               & &                &               \\
\midrule
1   &      100.00\% &      100.00\% & &       100.00\% &      100.00\% \\
8   &        5.24\% &       17.88\% & &        15.73\% &        4.40\% \\
9   &        3.79\% &       13.64\% & &        10.73\% &        3.30\% \\
41  &        0.09\% &        0.61\% & &         0.12\% &        1.10\% \\
42  &        0.09\% &        0.61\% & &         0.11\% &        1.10\% \\
\bottomrule
\end{tabular}
\end{table}

When looking at the ratios shown in \tabref{fig:kt:compare-ratios}, we observe a very interesting result.  First, for Kotlin files (on the right) we note that while the 2020 ratios are smaller but similar to the results we saw for the Java dataset, the 2014 ratios are much lower than the Java dataset's 2003 ratios.  This implies that Kotlin projects adopted method chaining much more recently, compared to Java projects that adopted it early on.

Second, we note the behavior of the Java files (on the left) in the Kotlin projects that seem to be exhibiting the opposite trend as the Kotlin files.  We suspect this might be due to some method chains moving from Java files over to their Kotlin replacements as people slowly replace existing Java code with newer Kotlin equivalents.

\findings {
    While chain length ratios are decreasing for Java source files in Kotlin projects, they are increasing for Kotlin files. This might imply developers moving chains from one file to another.
}

\begin{table}[th]
    \centering
    \caption{Groups of method chains (Kotlin files)}
    \label{tab:kt:kt:groups}
\newcommand{\oldtabcolsepAA}{\tabcolsep}
\renewcommand{\tabcolsep}{2pt}
\begin{tabular}{lrrrr}
\toprule
{} &                Short &              Long &          ExtLong \\
\midrule
Length range      &     1 < len. $\le$ 3 &      3 < len. < 8 &     8 $\le$ len. \\
\# chains in 2020 &  1,554,699 (89.97\%) &  160,373 (9.28\%) &  13,038 (0.75\%) \\
\# chains in 2014 &      1,303 (94.90\%) &       62 (4.52\%) &       8 (0.58\%) \\
\bottomrule
\end{tabular}
\renewcommand{\tabcolsep}{\oldtabcolsepAA}
\end{table}

When looking at the method chains grouped as shown in \tabref{tab:kt:kt:groups}, we see that Kotlin appears to have shorter overall method chains when compared to Java.  However, it appears that Kotlin has a higher ratio of long/extra long chains out of the three categories compared to Java. For example, in 2020, 9.28\% of chains are long chains in Kotlin while only 3.51\% of chains are long chains in Java, and 0.75\% of chains are extra long chains in Kotlin files while only 0.03\% are in Java.

\findings{
    Kotlin projects appear to have more long and extra long method chains compared to Java projects.
}

\begin{figure}[ht]
    \centering
    \includegraphics[width=0.49\linewidth]{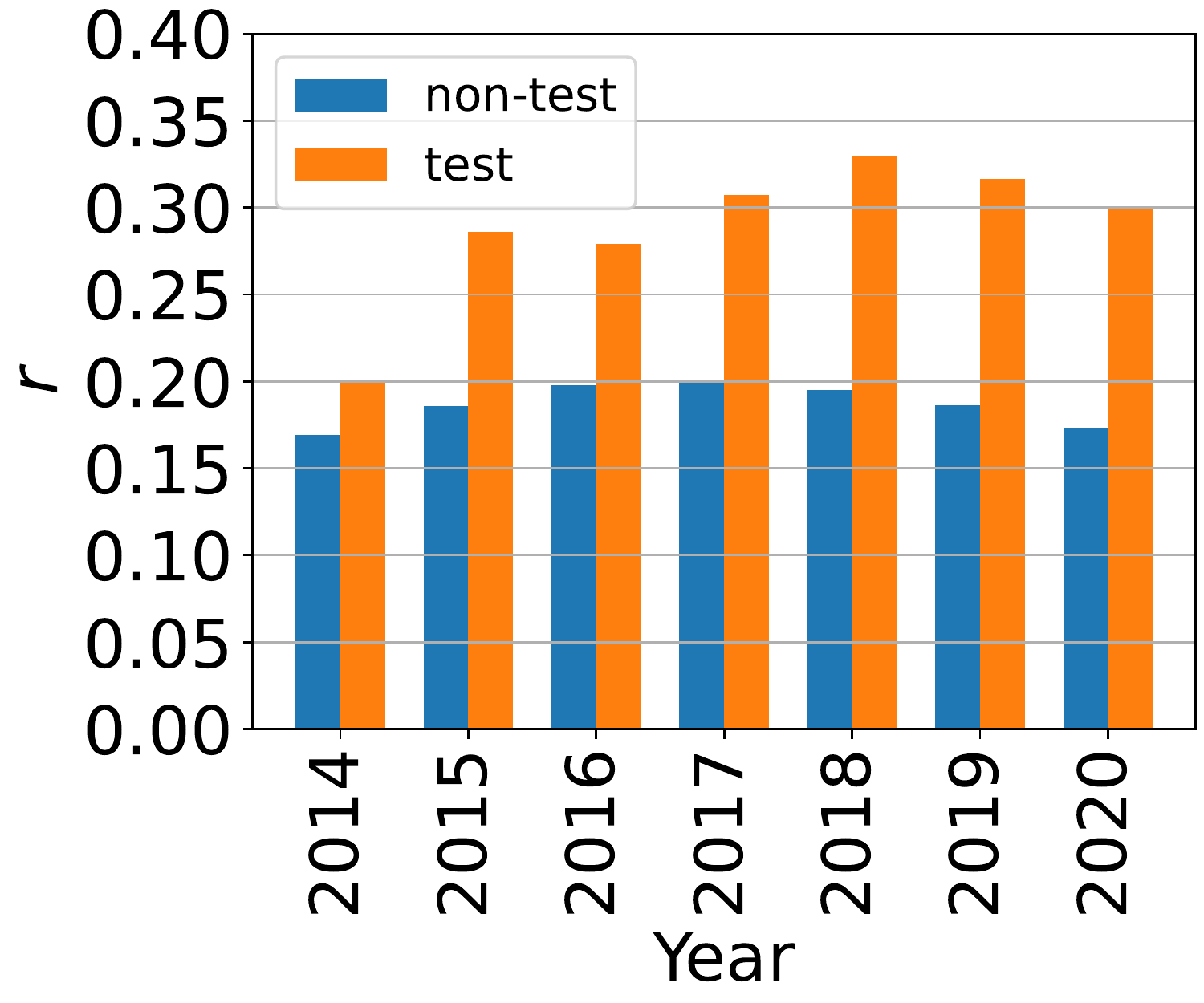}
    \includegraphics[width=0.49\linewidth]{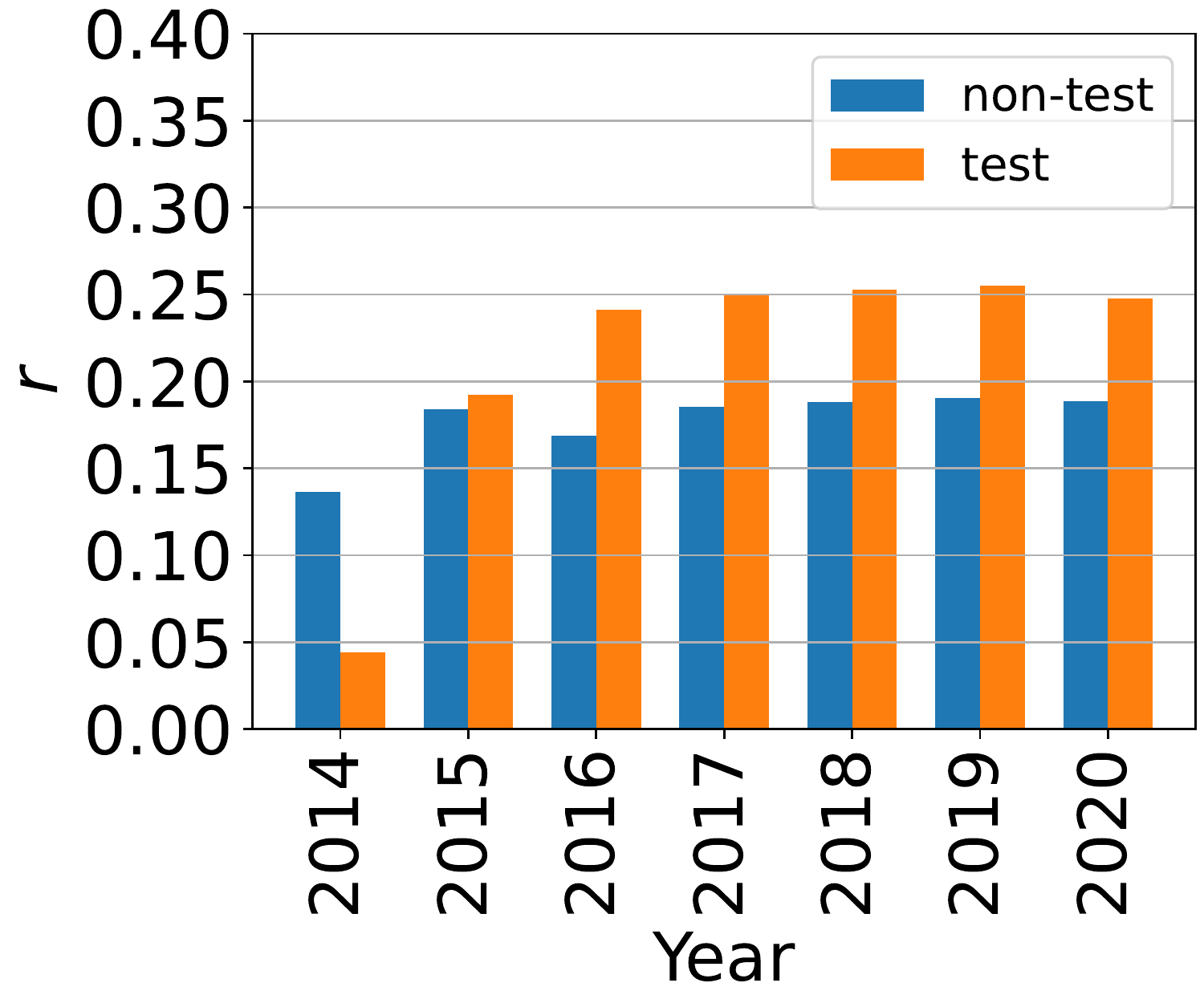}
    \caption{$r$ per year of non-testing vs. testing in Kotlin projects (Java files (left) and Kotlin files (right))}
    \label{fig:kt:r-test}
\end{figure}

Next we look at the distribution of chains in testing vs non-testing code.  \figref{fig:kt:r-test} shows the $r$ values per year for testing vs. non-testing code with Java files on the left and Kotlin files on the right.  For both languages, it appears starting from 2015, testing code has more method chains than non-testing code.  For Kotlin files, 2015 appears to be the point where the majority of method chains switched from being in non-testing code to testing code.

\begin{figure}[t]
    \centering
    \includegraphics[width=0.49\linewidth]{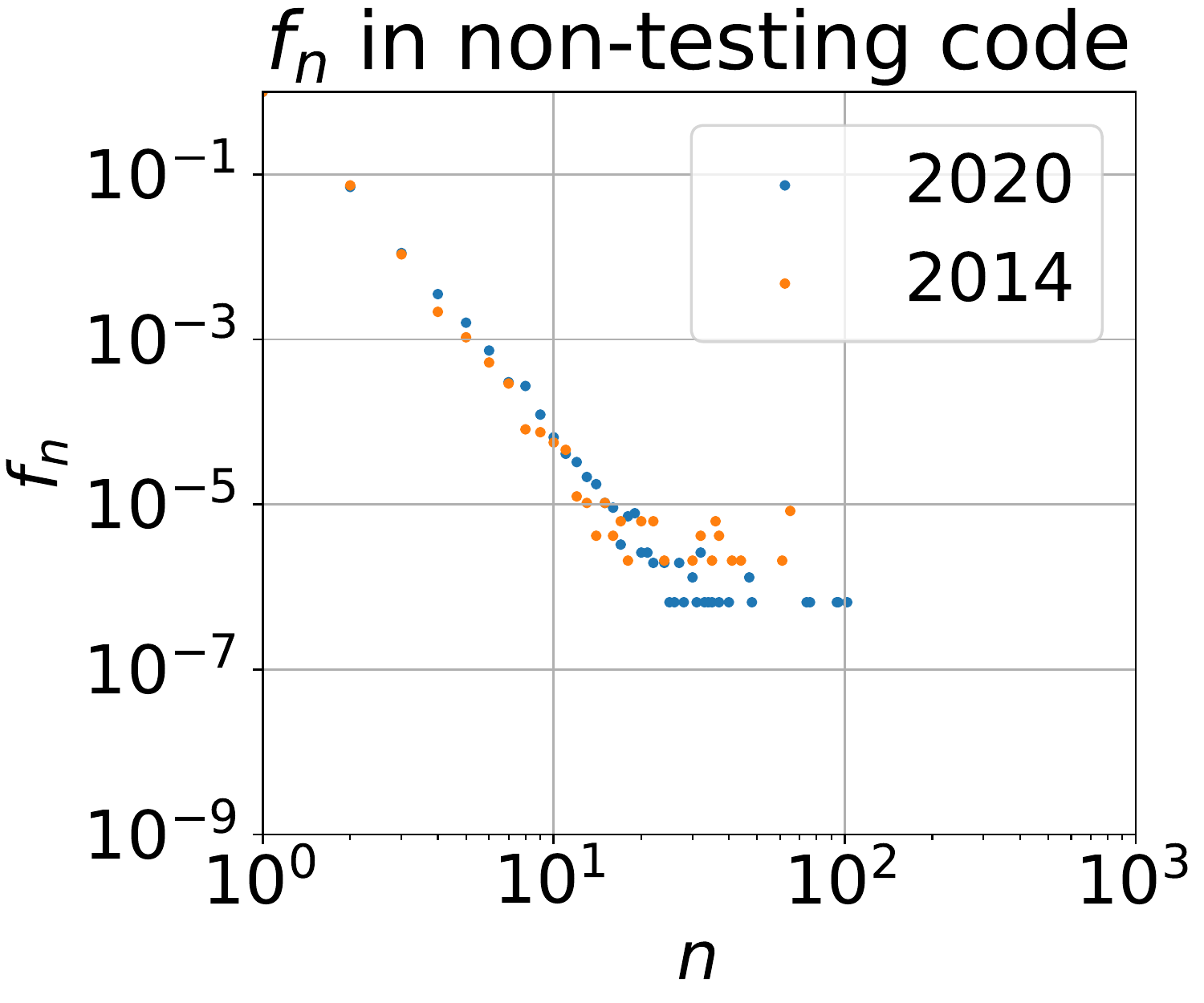}
    \includegraphics[width=0.49\linewidth]{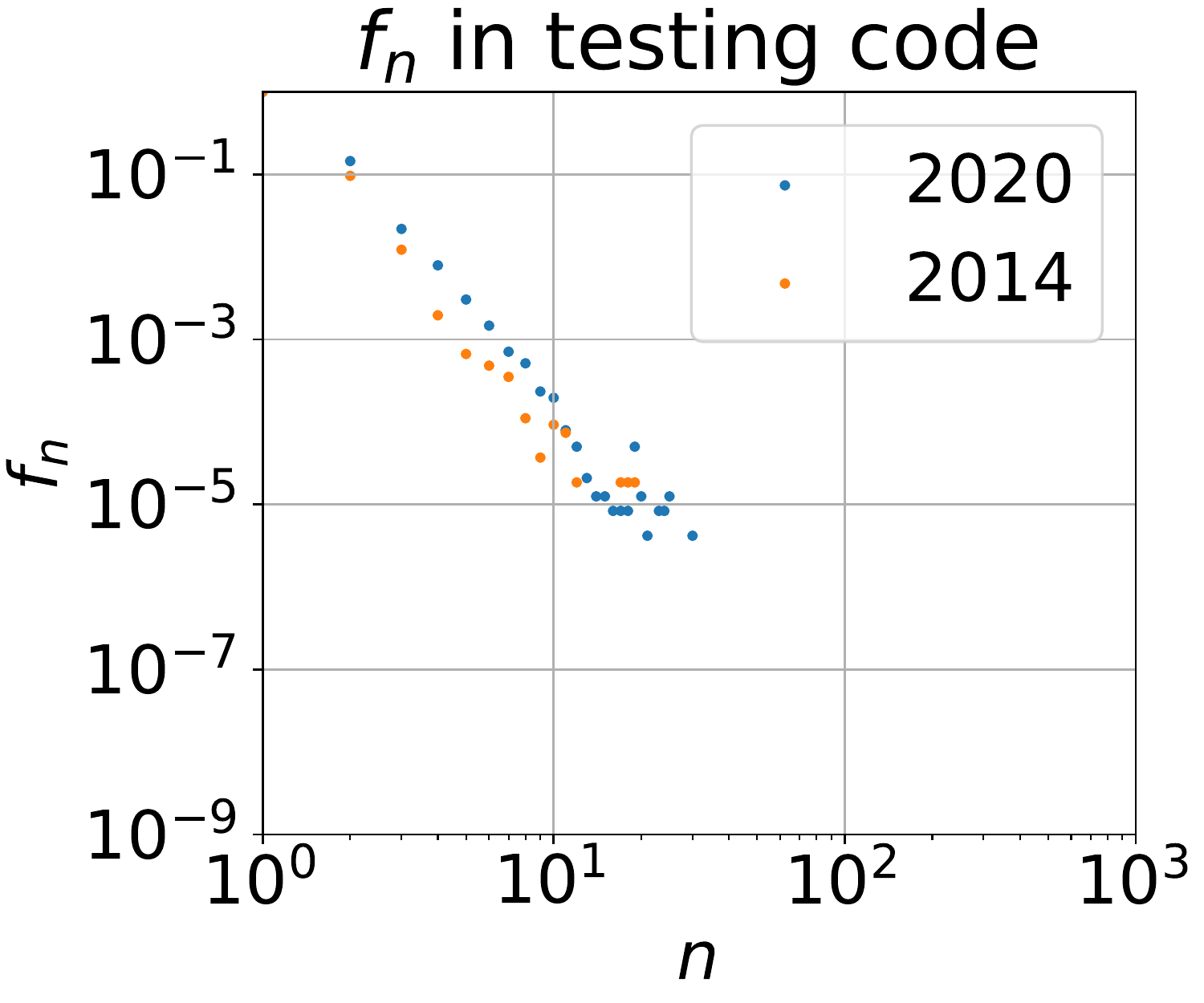}
    \caption{$f_n$ of non-testing vs testing code (Java files)}
    \label{fig:kt:Java:test}
\end{figure}

When we observe the scatter plots for the Java files, shown in \figref{fig:kt:Java:test}, we see that despite there being more overall chains in the testing code it appears that the lengths of those method chains are shorter than the method chains in non-testing code.

\begin{figure}[t]
    \centering
    \includegraphics[width=0.49\linewidth]{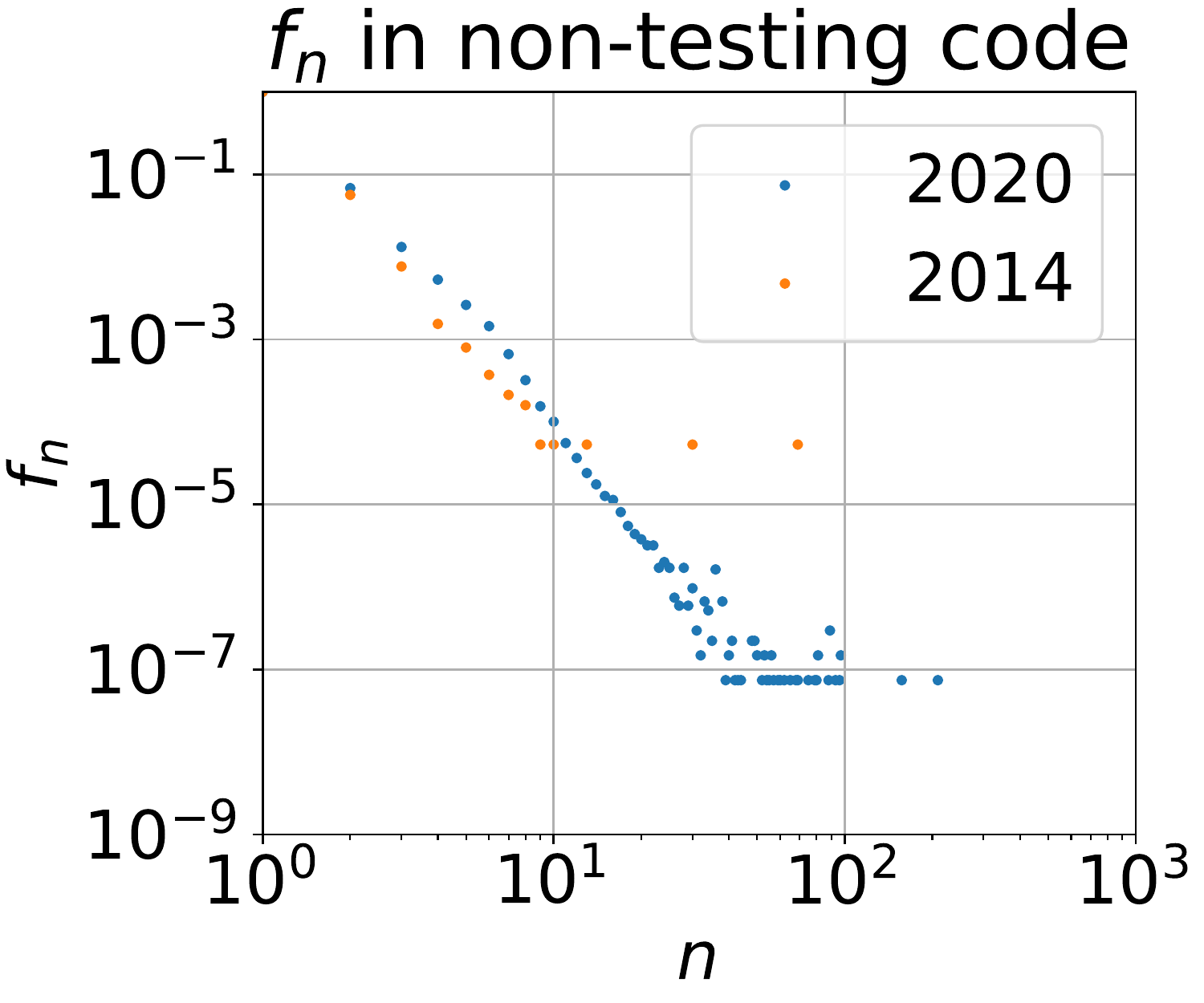}
    \includegraphics[width=0.49\linewidth]{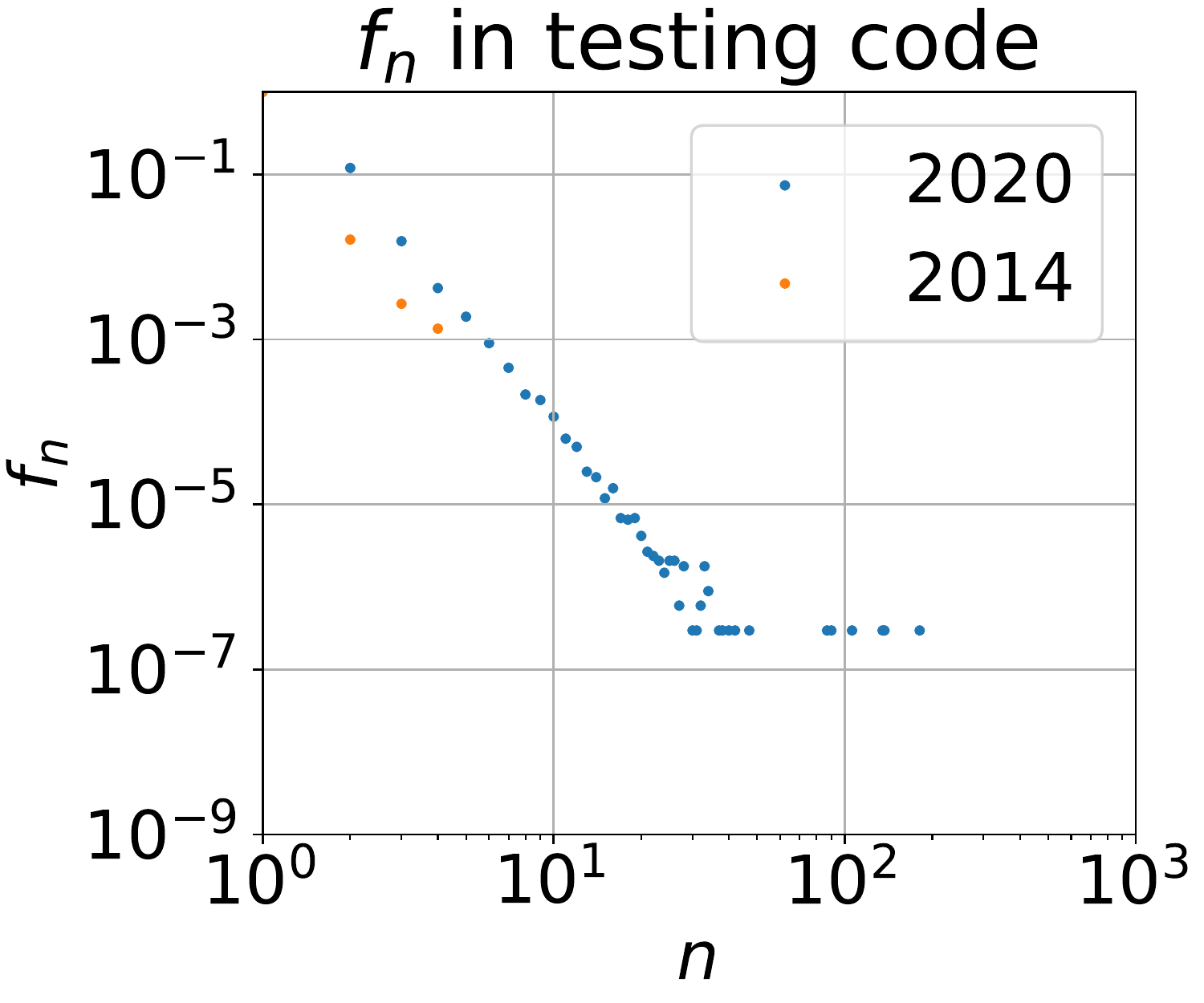}
    \caption{$f_n$ of non-testing vs testing code (Kotlin files)}
    \label{fig:kt:kt:test}
\end{figure}

However, when looking at the $f_n$ values for Kotlin files shown in \figref{fig:kt:kt:test}, both testing and non-testing code have similar chain lengths, and both have significantly more extra long chains in 2020 than in 2014.

When looking at the scatter plots for the Kotlin files, shown in \figref{fig:kt:kt:test}, there is a gap in the right tail of the testing code plot, suggesting that the longest chain lengths in 2020 are between 50 and 100, with a few outliers that are greater than 100 while the chain lengths for non-testing code, for the most part, steadily increase to around 100.  The fact that the $f_n$ values are generally lower for the non-testing scatter plot indicates that testing code has a higher ratio of chain lengths greater than 1 to chain lengths equal to 1. The scarcity of dots for 2014 in the testing scatter plot suggest that many chains in testing code were of certain repeated lengths.

\findings{
    Since 2015, testing code sees more method chaining among Java files and Kotlin files than non-testing code.  That being said, chain lengths tend to be longer in non-testing code than in testing code, overall.
}


We conclude by answering the research question: Like Java, Kotlin sees more method chaining in testing code than in non-testing code. Unlike Java, however, the use of method chaining in Kotlin is not increasing, but remaining relatively constant. Therefore, we can say that developers use method chaining in Java and Kotlin for similar purposes, but method chaining is not becoming more popular in Kotlin, as it is in Java.

\subsection{\ref*{rq-python}: Do Python programmers use method chaining in a way similar to Java or Kotlin programmers?}
\label{subsec:python}

Now that we looked at the use of method chains in Java and Kotlin, two JVM-based languages, we want to investigate one more language.  We chose to look at the Python language, a popular scripting language that is also not a JVM-based language.  Our hypothesis was that Python programmers will behave different compared to Java and Kotlin programmers.

\begin{figure}[t]
    \centering
    \includegraphics[width=0.49\linewidth]{figures/java-full/r-bar.pdf}
    \includegraphics[width=0.49\linewidth]{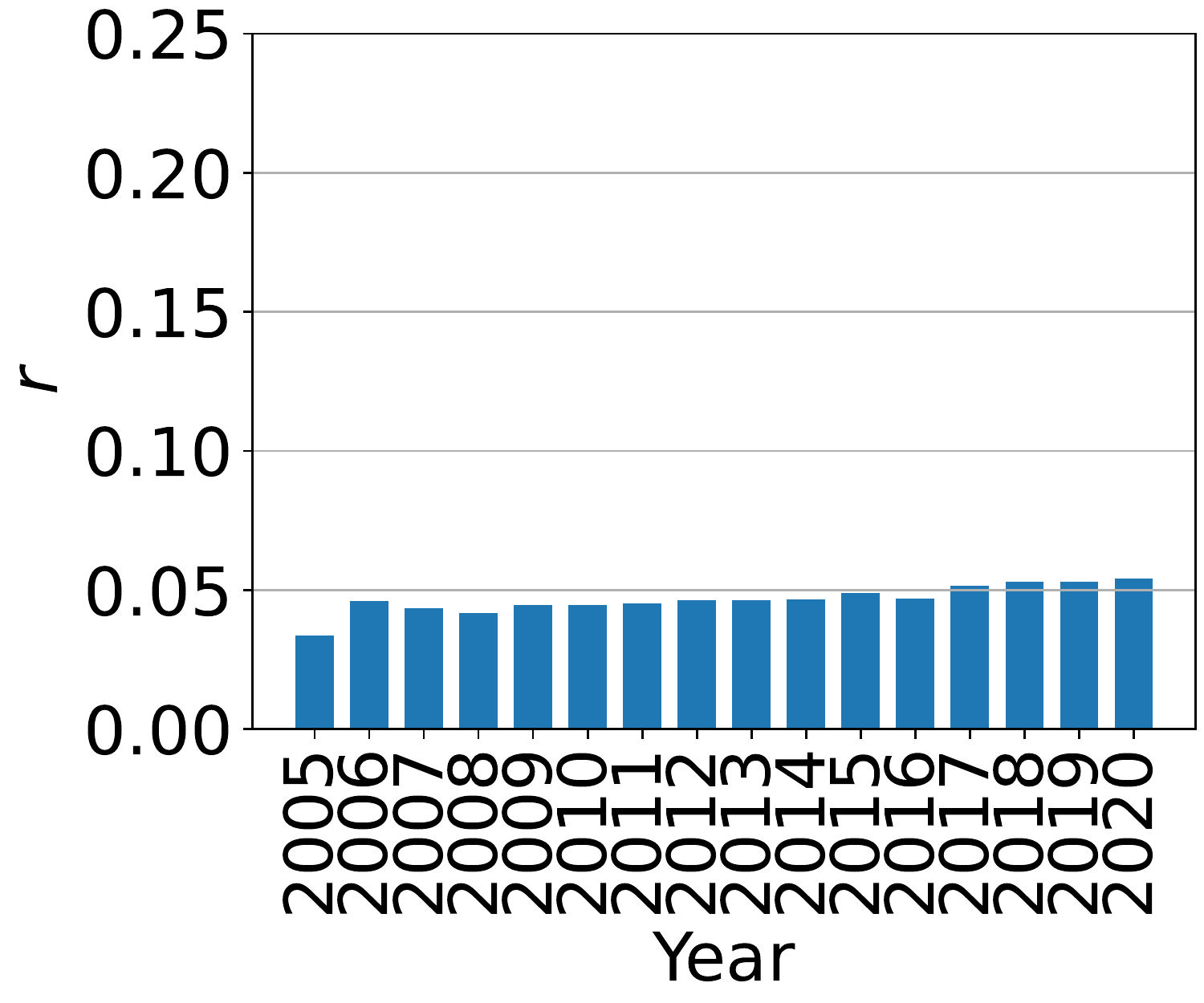}
    \caption{$r$ values (Java left, Python right)}
    \label{fig:py:scatter}
\end{figure}

The first thing we noticed was how different Python's $r$ values are compared to the other languages.  For example, you can see in \figref{fig:py:scatter} with Java on the left and Python on the right, Python's $r$ values are almost 3x lower than Java's.

\begin{figure}[ht]
    \hspace{2.25cm}\textbf{Java}\hspace{3.6cm}\textbf{Python}
    \begin{center}
        \includegraphics[width=0.49\linewidth]{figures/java-full/r-hist.pdf}
        \includegraphics[width=0.49\linewidth]{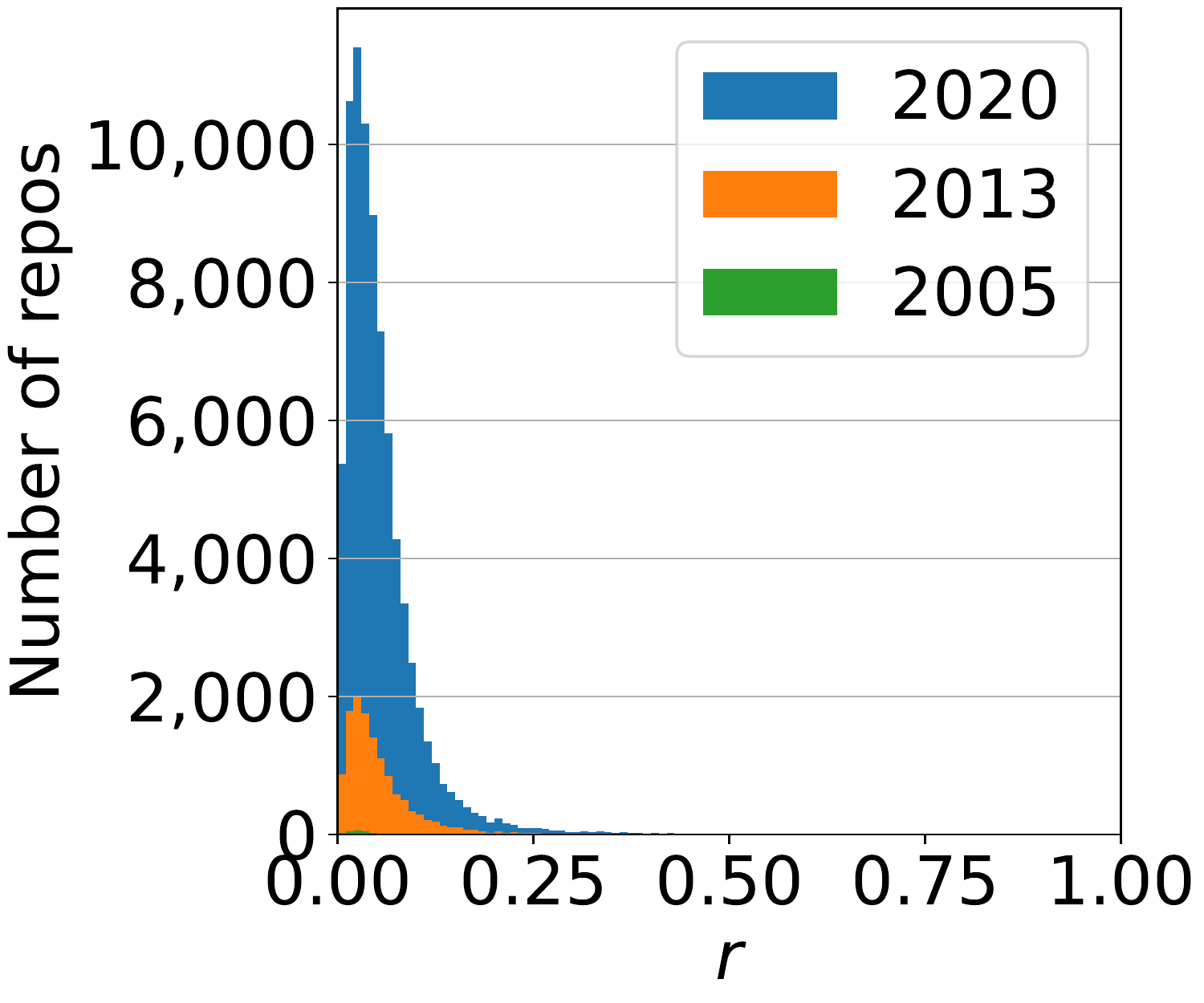}
    \end{center}
    \hspace{1cm}\textbf{Kotlin (Kotlin files)}\hspace{1.5cm}\textbf{Kotlin (Java files)}
    \begin{center}
    \includegraphics[width=0.49\linewidth]{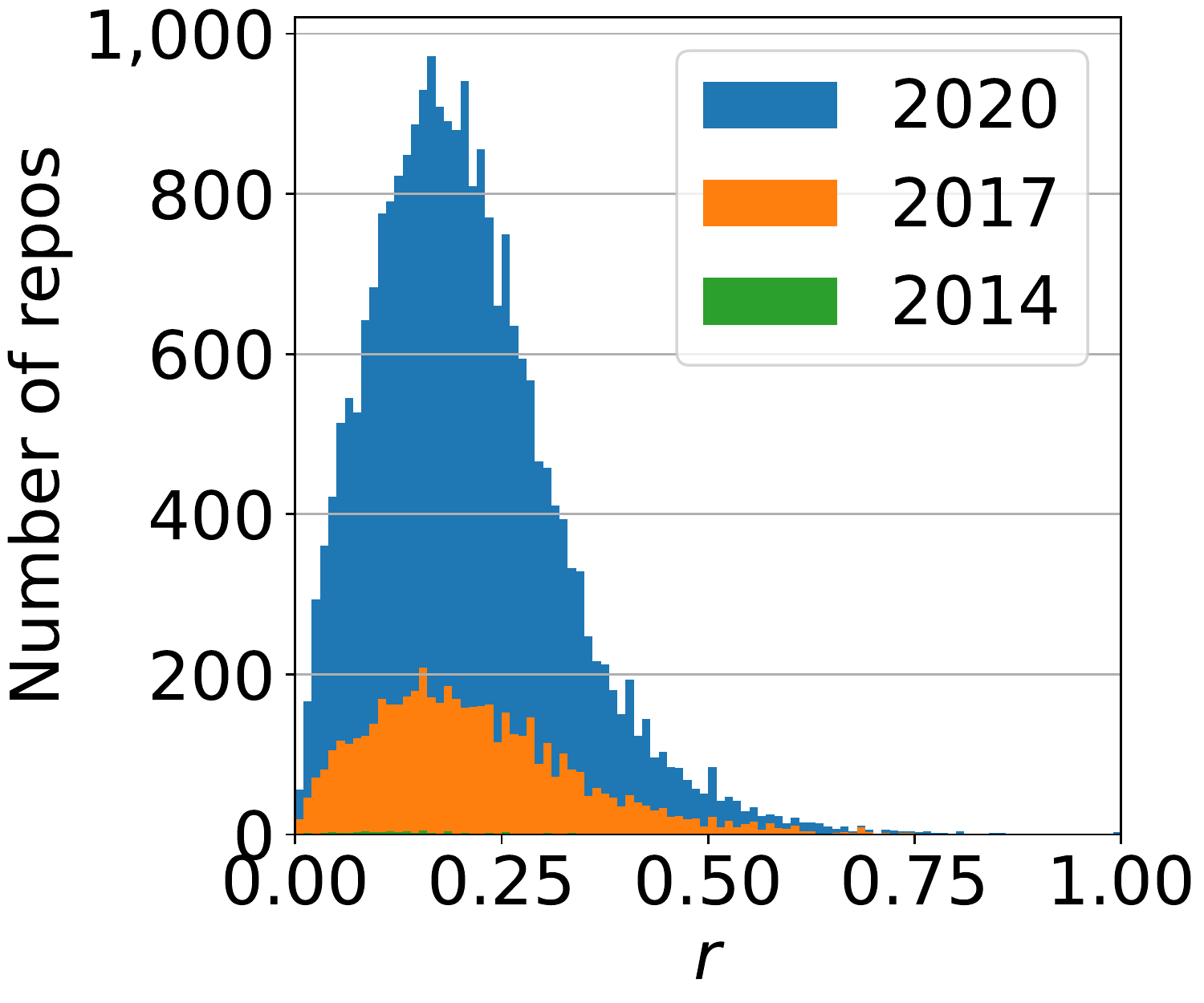}
    \includegraphics[width=0.49\linewidth]{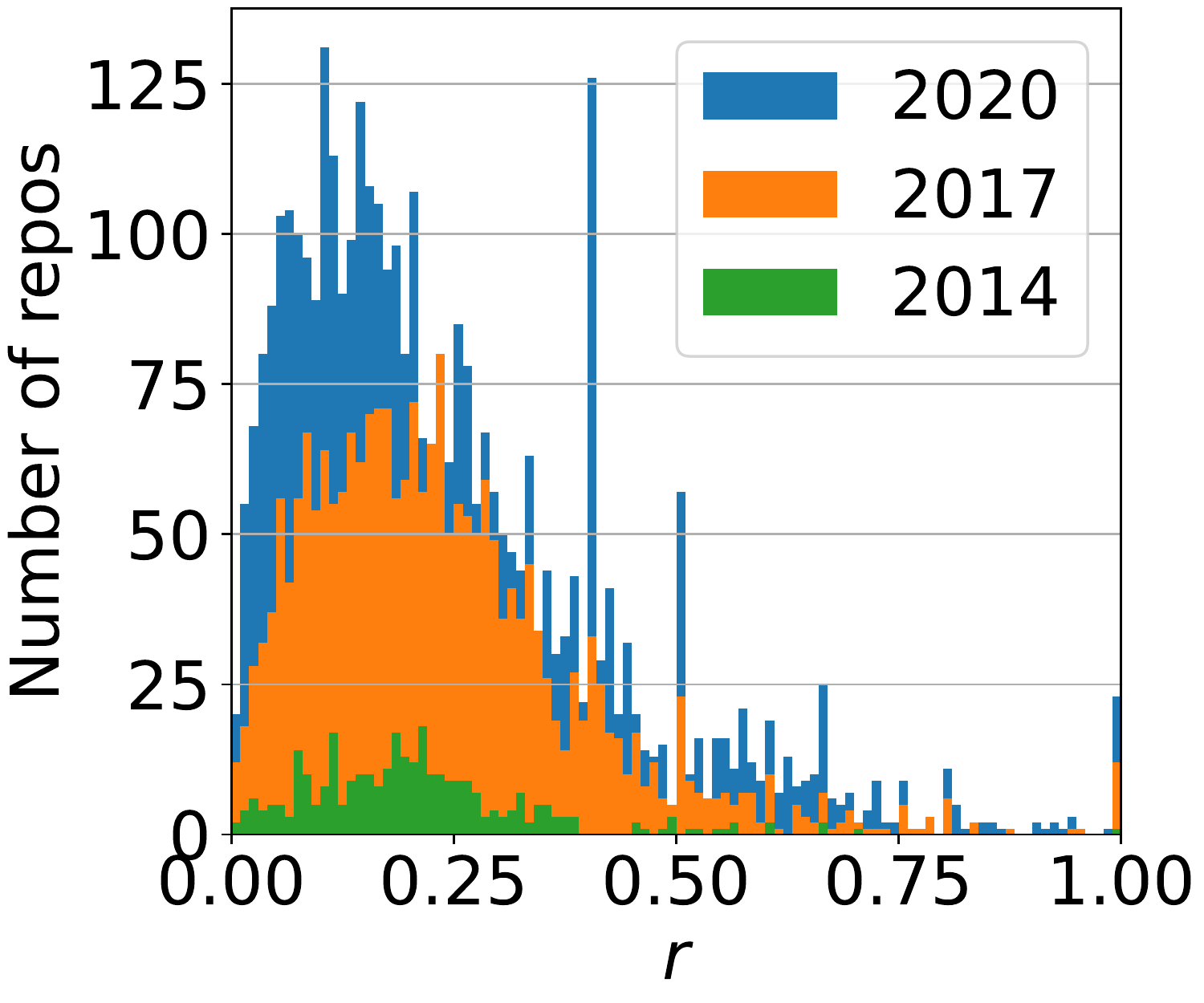}
    \end{center}
    \caption{Distribution of $r$ values per project across all studied languages}
    \label{fig:py:hist}
\end{figure}

In fact, we observe considerably less method chain use in Python than Java, which becomes even more obvious when we examine the histograms in \figref{fig:py:hist}, with Java on the left and Python on the right.  Keep in mind, there are 9k more Python projects in our dataset compared to the Java dataset.  While we do see similar shapes, as would be expected, the Python histogram is skewed so far left that it is starting to almost look like a line.

\findings{
    Python sees substantially less use of method chains when compared to either Java or Kotlin.
}

Similar to the other languages we also wanted to categorize the chains in Python into short, long, and extra long categories.  The results are shown in \tabref{tab:py:groups}.  Here we observe that Python is similar to the other languages, in that over time there are more long and extra long chains.

\begin{table}[t]
    \centering
    \caption{Groups of method chains (Python)}
    \label{tab:py:groups}
\newcommand{\oldtabcolsepA}{\tabcolsep}
\renewcommand{\tabcolsep}{2pt}
\begin{tabular}{lrrrr}
\toprule
{} &                Short &              Long &          ExtLong \\
\midrule
Length range      &     1 < len. $\le$ 3 &      3 < len. < 8 &     8 $\le$ len. \\
\# chains in 2020 &  3,889,936 (96.58\%) &  124,390 (3.09\%) &  13,397 (0.33\%) \\
\# chains in 2014 &    823,329 (97.88\%) &   15,704 (1.87\%) &   2,104 (0.25\%) \\
\# chains in 2005 &     19,024 (97.61\%) &      402 (2.06\%) &      64 (0.33\%) \\
\bottomrule
\end{tabular}
\renewcommand{\tabcolsep}{\oldtabcolsepA}
\end{table}

\begin{figure}[t]
    \centering
    \raisebox{-.5\height}{\includegraphics[width=0.5\linewidth]{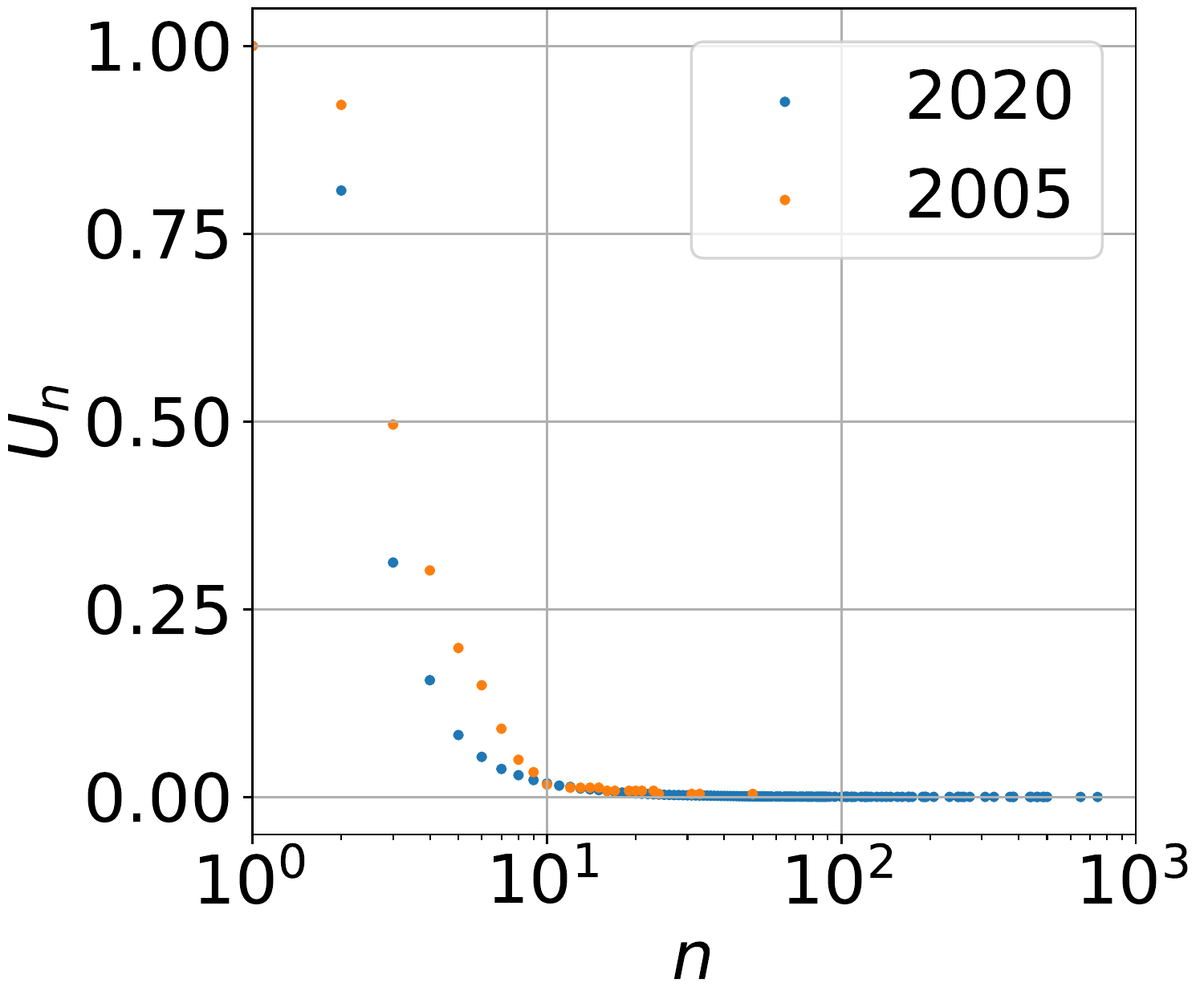}}
\newcommand{\oldtabcolsep}{\tabcolsep}
\renewcommand{\tabcolsep}{2pt}
\begin{tabular}{lrrrr}
\toprule
{} & $U_n$ in 2020 & $U_n$ in 2005 \\
$n$ &               &               \\
\midrule
1   &      100.00\% &      100.00\% \\
8   &        2.90\% &        4.96\% \\
9   &        2.25\% &        3.31\% \\
41  &        0.13\% &        0.41\% \\
42  &        0.13\% &        0.41\% \\
\bottomrule
\end{tabular}
\renewcommand{\tabcolsep}{\oldtabcolsep}
    \caption{Ratio of Python projects containing chains $\ge$ to $n$}
    \label{fig:py:ratios}
\end{figure}

When we observe the $U_n$ values for Python, shown in \figref{fig:py:ratios}, we see a very small percent of the projects have chains of length 8 or more.  This is a stark contrast with what we observed in Java and Kotlin where around 15--20\% of projects have chains of length 8 or more.  So not only does Python have substantially fewer chains, but the chains it does contain are typically much shorter.

\findings {
    Chains in Python tend to be much shorter than chains in Java or Kotlin.
}

\begin{figure}[t]
    \centering
    \includegraphics[width=0.49\linewidth]{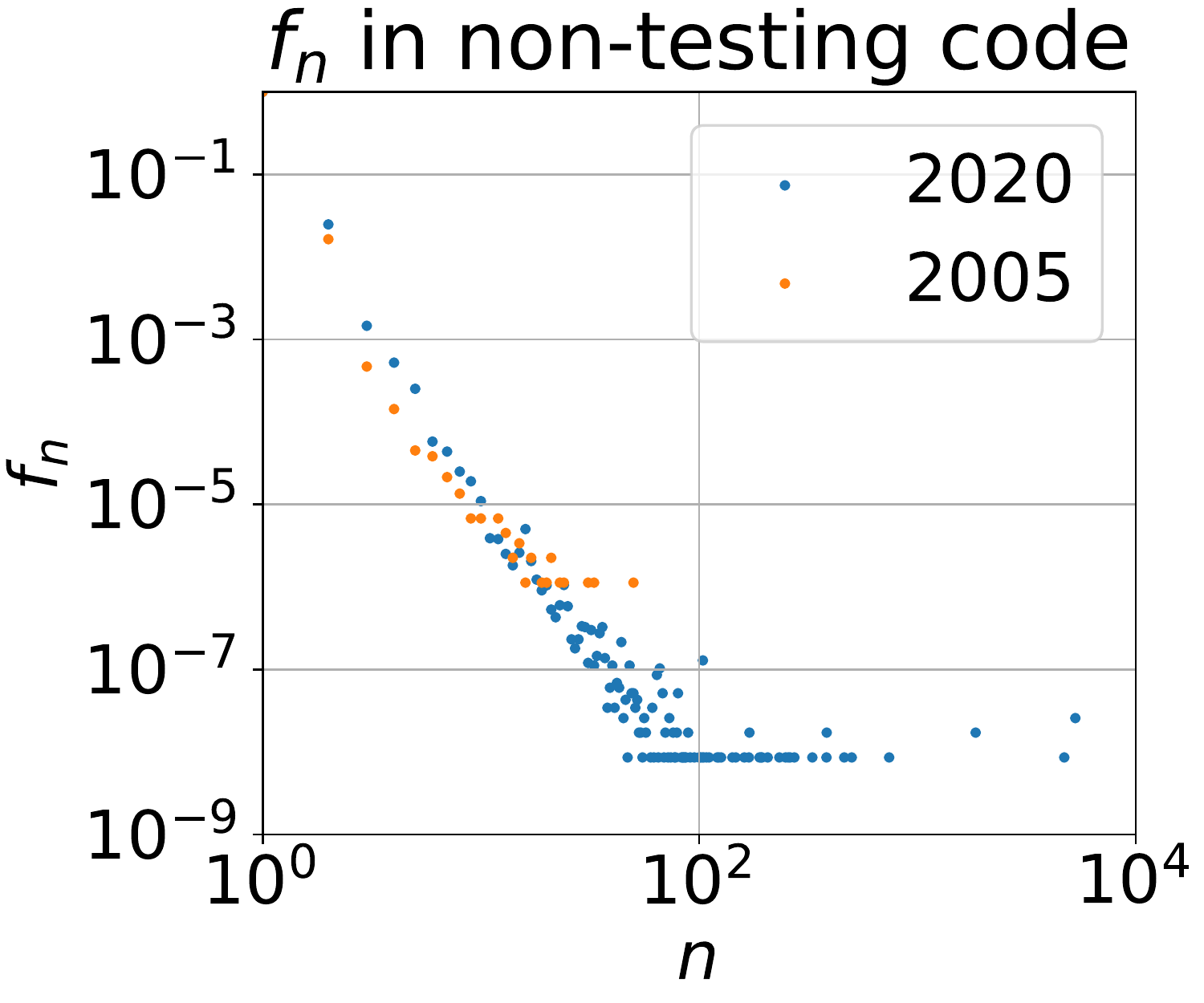}
    \includegraphics[width=0.49\linewidth]{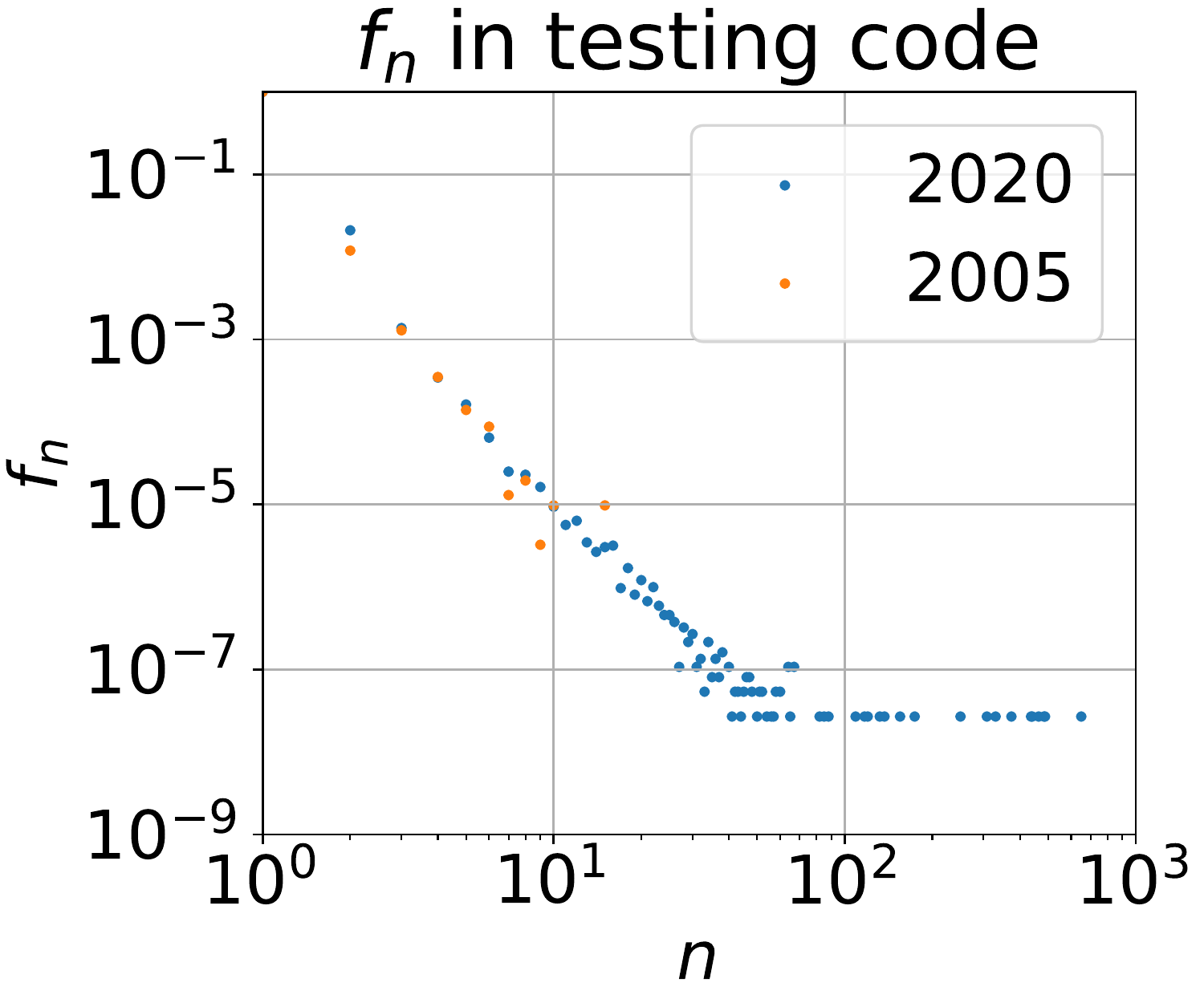}
    \caption{$f_n$ of non-testing vs testing code (Python)}
    \label{fig:py:test-scatter}
\end{figure}

\begin{figure}[t]
    \centering
    \includegraphics[width=0.5\linewidth]{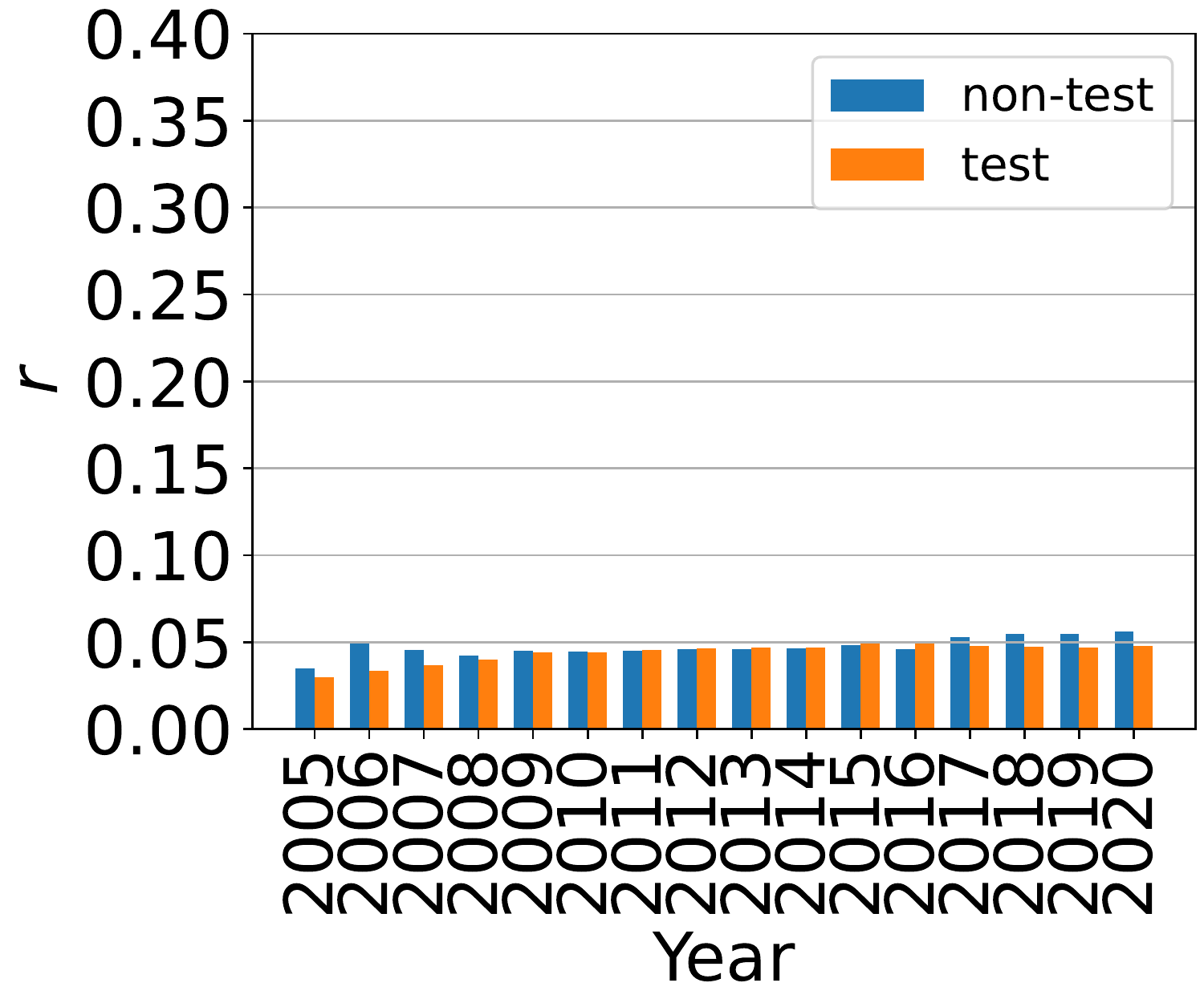}
    \caption{$r$ per year of non-testing code vs. testing Code (Python)}
    \label{fig:py:r-test}
\end{figure}

Finally, when looking at the behavior in testing vs non-testing code, we see additional differences when compared with the other two languages.  For both the Java and Python datasets, around 25\% of files were test files.  We can see in the scatter plot of \figref{fig:py:test-scatter} the $f_n$ values, while lower than Java, follow similar trends.  In \figref{fig:py:r-test} we can see the $r$ values and here we observe some differences to Java and Kotlin.  It appears that over time, Python is following a different trend and the use of chains is increasing in non-testing code while staying relatively flat for testing code.

\findings {
    Contrary to method chaining becoming more common in testing code for Java and Kotlin, for Python, method chaining is becoming more common in non-testing code.
}

We conclude by answering the research question: method chaining in Python occurs less frequently than in Java or Kotlin and chains in Python tend to be shorter than in Java or Kotlin. However, unlike Java and Kotlin, method chaining is more common in non-testing code than in testing code.

\subsection{\ref*{rq-extend}: Do we see support for the proposed language extensions from the prior study?}
\label{subsec:extend}

\citet{original-data} also investigated four possible language extensions for Java.  They then created a sample of 385 chains from their data (out of over 150 million), manually analyzed the sample, and estimated how often such a language extension could be applicable.  Here we take another look at the code patterns they identified, but don't sample the data and instead use an automated mining approach to see if we can observe support for the proposed language extensions.  Note that the data here is not deduplicated.

\subsubsection{NullExceptionAvoidance}

The first pattern was looking for method chains where the user checks the return value of one (or more) of the calls to ensure it is not \texttt{null} before continuing on with the chain.  The original study proposed a \textit{safe call} syntax, similar to what Kotlin provides:

\begin{lstlisting}[numbers=none]
    m1()?.m2().m3()
\end{lstlisting}

\noindent where the whole expression evaluates to \texttt{null} (and \texttt{m2} and \texttt{m3} are not called) if the call to \texttt{m1} returns \texttt{null}.  Since we already have a very large dataset with almost 500k Kotlin projects, we investigated how frequently such a feature is used in the full dataset (not just the 26k projects we mined for method chains).

In total, we found that 257,114 projects (51.5\%) used at least one safe call and there were a total of 4,141,922 safe calls.  On average, projects contained 8.3 safe calls spanning 2.5 files.  If we focus just on our 26k studied projects, there were 355,132 safe calls in 22,041 projects (83.9\%), averaging 21.9 safe calls in 6.2 files.
This shows this feature is widely used in Kotlin and supports the conclusion that Java could possibly benefit from its addition.

\subsubsection{RepeatedReceiver}

The second pattern was looking for when two separate method chains that appear as neighboring statements both have their original call on the same receiver:

\begin{lstlisting}[numbers=none]
    o.m1();
    o.m2();
\end{lstlisting}

In such a case, these could be chained together if the API being called was modified to support chaining (e.g., returning the \texttt{this} object from \texttt{m1()}).
We queried the Java Original dataset and found 6,515,461 instances of this pattern.  This represents about 19.07\% of the method chains in that dataset.

The original paper had estimated there were 8.57\% with a 99\% confidence interval of 4.9--12.24\%.
Thus the actual value appears to be higher and outside their confidence interval.  There seems to be support for possibly refactoring APIs into a more fluent style.

\subsubsection{DownCast}

The third pattern they suggested looked for a chain that contains a cast operation:

\begin{lstlisting}[numbers=none]
    ((C)m1()).m2().m3();
\end{lstlisting}

The idea was that frequently this kind of down-cast is difficult to read and many developers split the chain to store the cast result into a local.  An improved solution could be to support a method that performs the downcast for you, e.g.:

\begin{lstlisting}[numbers=none]
    m1().asC().m2().m3();
\end{lstlisting}

This would lead to more readable code and enable chaining the methods together without splitting to store in a local.

In total, we found 16,136 (0.047\%) chains matched this pattern and contained a cast in the Java Original dataset.
This is considerably smaller than their estimate of 1.56\%, but they were unable to compute a 99\% confidence interval so this could fall within such margins.

\subsubsection{ConditionalExecution}

The final pattern looked for method chains that were guarded with an if conditional:

\begin{lstlisting}[numbers=none]
    if (o.m1())
        o.m2()
\end{lstlisting}

\noindent and recommended a method that takes a lambda (the body of the conditional) and runs it if the condition is true.  So in this example, it might look something like \lstinline|o.ifM1(x->x.m2())|.

Again, when looking at the Java Original dataset we were able to mine for this pattern and found a total of 32,652 instances (0.0956\%).
The original paper had estimated there were 2.34\% with a 99\% confidence interval of 0.36--4.32\%.  Thus the actual value appears to be lower and outside their confidence interval.

\findings{
    While we were able to more thoroughly mine the method chains to see if the suggested patterns occurred, we found in half the cases they occur very infrequently, often lower than the 99\% confidence interval suggested by the prior study.  We did however find strong support for the NullExceptionAvoidance language feature recommended by the prior study.
}

We conclude by answering the research question: we see support for half the language extensions previously proposed.

\section{Threats to Validity}
\label{sec:threats}

Some possible threats to the internal validity of this study are that we do not know what kinds of projects are in the dataset, and thus there may be toy or educational projects included that could potentially skew the results (for example if someone was practicing how to use fluent APIs, all of their code might have a lot of chains).  Filtering by star counts helps avoid some projects, but may not catch them all.

There may also be partial clones data in the dataset.  We were only able to easily remove exact duplicates, aka type-1 clones.  There may be additional clones in the dataset and depending on the quantity it could be skewing the results.

For RQ2, we looked at commonly used libraries in extra long chains.  This analysis relied on a complicated Boa query attempting to infer types in the code.  While we believe this query is sound, it is possible the type inference was not complete and the list of popular libraries would be affected.

There are some threats to external validity.  Similar to the prior study, the main filter applied to select projects was star counts.  Thus, it is possible the trends we observe might be different for less popular projects.  And while this study expanded the original to look at two additional languages, the results we see may not generalize to other languages.

\section{Related Works}
\label{sec:related}

Many researchers have studied the use of Java language features~\cite{dyer2014mining}, especially the use of lambdas~\cite{Mazinanian17,nielebock19,petrulio21,Zheng21,Lucas19}, but few have looked at the use of normal methods (not lambdas) or method chaining in particular. Tanaka et al.~\cite{tanaka2019study} did look at method chains, but from the viewpoint of functional idioms in Java.

B\"{o}rstler and Pacch~\cite{borstler16} performed a study on the perceived readability of code that included method chains.  Their results indicated there was no significant relationship observed, so it is not clear if method chains increase or decrease readability and perhaps when looking at how they impact comprehension, readability might not be the right metric.  We did not investigate method chains impact on comprehension but rather just observed if people use method chaining in the wild.

Kasraee and Lin~\cite{Kasraee946689} performed an eye tracking study with participants reading code containing method chains.  Their results indicate that code without method chains may be slightly more readable.  If that is true, then the observations of our study indicate a lot of code could be made more readable by converting it to a non-chained form.

Grechanik et al.~\cite{10.1145/1852786.1852801} performed an empirical evaluation on Java projects.  They note that most methods have either one or zero arguments.  This result could impact how frequently fluent APIs get created, as some of the use cases of such APIs rely either on passing values in each call in the chain.  It might be interesting in the future to see how method chain arguments are typically used.

Kabanov and Raudj\"{a}rv~\cite{10.1145/1411732.1411758} came to the conclusion that a combination of the fluent interface idiom, static functions, metadata, and closures is a better coding practice than method chaining.  Their focus is on embedded DSLs for Java and they do not empirically investigate how developers have used method chaining in the past to drive their decisions.

\section{Conclusion and Future Work}
\label{sec:conclusion}

The trends of method chaining were not well understood, outside of Java.  A prior study looked at the use of chaining in 2.7k Java projects.  It was not clear if those results generalized to more Java projects or other languages.  In this work, we first replicated their prior results then generalized them to a larger Java dataset and observed similar trends: the use of method chains is popular and increasing.  We then investigated if those trends held for two other languages: Kotlin and Python.  While some of the trends were similar in Kotlin, it turns out the use of method chains in Python was quite different.  In Python, method chains are used considerably less frequently and when they are, the chains are generally shorter.  Additionally, while Java and Kotlin see more use of method chains in testing code, Python saw the opposite: more use in non-testing code.

Finding 11 showed that Python developers use substantially fewer method chains, compared to Java. The actionable result here is for API designers, as Python API designers may wish to avoid fluent APIs as it seems Python developers tend to not use chains as much.  Conversely, Java developers seem comfortable using chains, and API designers for Java may wish to employ fluent designs.

Now that we know developers use chains, some future work may investigate what common patterns appear as chains and see if a more succinct API can be developed.  Or perhaps IDE developers could provide code snippets for commonly occurring chains.  Because we know that a large portion of extra long chains (1/3) come from only five libraries in Java (Finding 6), this feature would be especially helpful for programmers who have to write very long chains. Further research can be done into what method chain templates are most useful for programmers.

We also envision a followup study that is more qualitative in nature to try and determine why there are differences among the languages: is it a lack of available fluent APIs, or do developers actively avoid chains in Python? Perhaps the idiomatic style of Python discourages writing chains?  An extension of the previous eye tracking study~\cite{Kasraee946689} to Python might show interesting differences.

\section{Data Availability}
The Boa queries, their outputs, and all processing scripts are available in a replication package~\cite{replicationpkg} on Zenodo.

\section*{Acknowledgements}
This work was partially funded by the UNL First Year Research Experience (FYRE) program.  We thank Tomoki Nakamaru for many clarifications and sharing scripts from the original study.

\printbibliography

@misc{original-errata,
  title={{On "An Empirical Study of Method Chaining in Java"}},
  author={Nakamaru, Tomoki},
  year={2022},
  howpublished={\url{https://tomokinakamaru.github.io/msr2020/}},
}

@dataset{original-data,
  author={Nakamaru, Tomoki and Matsunaga, Tomomasa and Yamazaki, Tetsuro and Akiyama, Soramichi and Chiba, Shigeru},
  title={Data - An Empirical Study of Method Chaining in {Java}},
  month=mar,
  year=2020,
  organization={Zenodo},
  publisher={Zenodo},
  version={1.0.0},
  doi={10.5281/zenodo.3697939}
}

@dataset{replicationpkg,
  author={Blinded.},
  title={{Replication package for ''Method Chaining Redux: An Empirical Study of Method Chaining in Java, Kotlin, and Python''}},
  month=mar,
  year=2022,
  organization={Zenodo},
  publisher={Zenodo},
  version={1.0.0},
  doi={10.5281/zenodo.6332602}
}

@inproceedings{original-study,
  author={Nakamaru, Tomoki and Matsunaga, Tomomasa and Yamazaki, Tetsuro and Akiyama, Soramichi and Chiba, Shigeru},
  title={An Empirical Study of Method Chaining in {Java}},
  year={2020},
  isbn={9781450375177},
  publisher={Association for Computing Machinery},
  address={New York, NY, USA},
  doi={10.1145/3379597.3387441},
  booktitle={Proceedings of the 17th International Conference on Mining Software Repositories},
  pages={93–102},
  numpages={10},
  location={Seoul, Republic of Korea},
  series={MSR '20},
}

@inproceedings{sha,
  title={Software heritage: Why and how to preserve software source code},
  author={Di Cosmo, Roberto and Zacchiroli, Stefano},
  booktitle={14th International Conference on Digital Preservation},
  publisher={{State Library of Victoria}},
  address={Melbourne, Australia},
  series={iPRES 2017},
  pages={1--10},
  year={2017}
}

@inproceedings{boa,
  author = {Dyer, Robert and Nguyen, Hoan Anh and Rajan, Hridesh and Nguyen, Tien N.},
  title = {{Boa}: A Language and Infrastructure for Analyzing Ultra-Large-Scale Software Repositories},
  year = {2013},
  isbn = {9781467330763},
  publisher = {IEEE Press},
  address={Piscataway, New Jersey},
  booktitle = {Proceedings of the 2013 International Conference on Software Engineering},
  pages = {422–431},
  numpages = {10},
  location = {San Francisco, CA, USA},
  series = {ICSE '13}
}

@misc{boa-website,
  author={Rajan, Hridesh and Nguyen, Tien N. and Dyer, Robert and Nguyen, Hoan Anh},
  title={{Boa}--Mining Ultra-Large-Scale Software Repositories website},
  howpublished={\url{http://boa.cs.iastate.edu/boa/}},
  year={2021},
}

@article{tanaka2019study,
  title={A Study on the Current Status of Functional Idioms in {Java}},
  author={Tanaka, Hiroto and Matsumoto, Shinsuke and Kusumoto, Shinji},
  journal={IEICE Transactions on Information and Systems},
  volume={102},
  number={12},
  pages={2414--2422},
  year={2019},
  publisher={The Institute of Electronics, Information and Communication Engineers},
}

@inproceedings{dyer2014mining,
  author = {Dyer, Robert and Rajan, Hridesh and Nguyen, Hoan Anh and Nguyen, Tien N.},
  title = {Mining Billions of {AST} Nodes to Study Actual and Potential Usage of {Java} Language Features},
  year = {2014},
  isbn = {9781450327565},
  publisher = {Association for Computing Machinery},
  address = {New York, NY, USA},
  doi = {10.1145/2568225.2568295},
  booktitle = {Proceedings of the 36th International Conference on Software Engineering},
  pages = {779–790},
  numpages = {12},
  location = {Hyderabad, India},
  series = {ICSE 2014}
}

@inproceedings{10.1145/1852786.1852801,
  author={Grechanik, Mark and McMillan, Collin and DeFerrari, Luca and Comi, Marco and Crespi, Stefano and Poshyvanyk, Denys and Fu, Chen and Xie, Qing and Ghezzi, Carlo},
  title={An Empirical Investigation into a Large-Scale {Java} Open Source Code Repository},
  year={2010},
  isbn={9781450300391},
  publisher={Association for Computing Machinery},
  address={New York, NY, USA},
  doi={10.1145/1852786.1852801},
  booktitle={Proceedings of the 2010 ACM-IEEE International Symposium on   Empirical Software Engineering and Measurement},
  articleno={11},
  numpages={10},
  location={Bolzano-Bozen, Italy},
  series={ESEM '10},
}

@inproceedings{10.1145/1411732.1411758,
  author={Kabanov, Jevgeni and Raudj\"{a}rv, Rein},
  title={Embedded Typesafe Domain Specific Languages for {Java}},
  year={2008},
  isbn={9781605582238},
  publisher={Association for Computing Machinery},
  address={New York, NY, USA},
  doi={10.1145/1411732.1411758},
  booktitle={Proceedings of the 6th International Symposium on Principles and Practice of Programming in Java},
  pages={189–197},
  numpages={9},
  location={Modena, Italy},
  series={PPPJ '08},
}

@misc{fluent,
  author={Fowler, Martin},
  title={{FluentInterface}},
  year={2005},
  howpublished={\url{https://martinfowler.com/bliki/FluentInterface.html}},
}

@misc{so-conflict,
  author={Kajaste, Ilari},
  year={2009},
  title={{Stack Overflow}: Method chaining - why is it a good practice, or not?},
  howpublished={\url{https://stackoverflow.com/questions/1103985/method-chaining-why-is-it-a-good-practice-or-not}},
}

@article{demeter,
  author={Lieberherr, K.J. and Holland, I.M.},
  journal={IEEE Software},
  title={Assuring good style for object-oriented programs},
  year={1989},
  volume={6},
  number={5},
  pages={38-48},
  doi={10.1109/52.35588},
}

@book{gamma1994design,
  author = {Gamma, Erich and Helm, Richard and Johnson, Ralph and Vlissides, John M.},
  title = {Design Patterns: Elements of Reusable Object-Oriented Software},
  year = {1994},
  publisher = {Addison-Wesley Professional},
  address = {Boston, MA},
  isbn = {0201633612},
}

@book{Fowler10,
  address = {Upper Saddle River, NJ},
  author = {Fowler, Martin},
  publisher = {Addison-Wesley},
  title = {Domain-Specific Languages},
  year = 2010,
}

@misc{linq,
  title={Language Integrated Query ({LINQ}) ({C\#})},
  author={Microsoft},
  year={2007},
  howpublished={\url{https://docs.microsoft.com/en-us/dotnet/csharp/programming-guide/concepts/linq/}},
}

@misc{fluentbad,
  title={Fluent Interfaces Are Bad for Maintainability},
  author={Bugayenko, Yegor},
  year={2018},
  howpublished={\url{https://www.yegor256.com/2018/03/13/fluent-interfaces.html}},
}

@inproceedings{lukas10,
  author = {Rytz, Lukas and Odersky, Martin},
  title = {Named and Default Arguments for Polymorphic Object-Oriented Languages: A Discussion on the Design Implemented in the {Scala} Language},
  year = {2010},
  publisher = {Association for Computing Machinery},
  address = {New York, NY, USA},
  doi = {10.1145/1774088.1774529},
  booktitle = {Proceedings of the 2010 ACM Symposium on Applied Computing},
  pages = {2090–2095},
  numpages = {6},
  location = {Sierre, Switzerland},
  series = {SAC '10},
}

@mastersthesis{Kasraee946689,
   author = {Kasraee, Pezhman and Lin, Chong},
   school = {Blekinge Institute of Technology, Department of Software Engineering},
   pages = {105},
   title = {Readability of Method Chains: A Controlled Experiment with Eye Tracking Approach},
   year = {2016},
}

@misc{tiobe,
  title={{TIOBE} Index for {January} 2023},
  author = {{TIOBE Software BV}},
  year = {2023},
  month = {1},
  url = {https://www.tiobe.com/tiobe-index/},
}

@ARTICLE{borstler16,
  author={B\"{o}rstler, J\"{u}rgen and Paech, Barbara},
  journal={IEEE Transactions on Software Engineering},
  title={The Role of Method Chains and Comments in Software Readability and Comprehension—An Experiment},
  year={2016},
  volume={42},
  number={9},
  pages={886-898},
  doi={10.1109/TSE.2016.2527791},
}

@article{petrulio21,
  title={The indolent lambdification of {Java} Understanding the support for lambda expressions in the {Java} ecosystem},
  author={Petrulio, Fernando and Sawant, Anand Ashok and Bacchelli, Alberto},
  journal={Empirical Software Engineering},
  volume={26},
  number={6},
  pages={134:1--134:36},
  year={2021},
  publisher={Springer},
  address={Netherlands},
}

@article{nielebock19,
  author = {Nielebock, Sebastian and Heum\"{u}ller, Robert and Ortmeier, Frank},
  title = {Programmers Do Not Favor Lambda Expressions for Concurrent Object-Oriented Code},
  year = {2019},
  publisher = {Springer},
  address = {USA},
  volume = {24},
  number = {1},
  issn = {1382-3256},
  doi = {10.1007/s10664-018-9622-9},
  journal = {Empirical Software Engineering},
  month = {2},
  pages = {103-138},
  numpages = {36},
}

@inproceedings{Lucas19,
  author={Lucas, Walter and Bonif\'{a}cio, Rodrigo and Canedo, Edna Dias and Marc\'{i}lio, Diego and Lima, Fernanda},
  title={Does the Introduction of Lambda Expressions Improve the Comprehension of {Java} Programs?},
  year={2019},
  isbn={9781450376518},
  publisher={Association for Computing Machinery},
  address={New York, NY, USA},
  doi={10.1145/3350768.3350791},
  booktitle={Proceedings of the XXXIII Brazilian Symposium on Software Engineering},
  pages={187-196},
  numpages={10},
  location={Salvador, Brazil},
  series={SBES 2019}
}

@Article{lopes17:_dejav,
  month = {10},
  year = {2017},
  doi = {10.1145/3133908},
  eid = {84},
  numpages = {28},
  number = {OOPSLA},
  volume = {1},
  date = {2017-10},
  journal = {Proceedings of the {ACM} on Programming Languages},
  title = {{D\'ej\`aVu}: A map of Code Duplication on {GitHub}},
  author = {Lopes, Cristina V. and Maj, Petr and Martins, Pedro and Saini, Vaibhav and Yang, Di and Zitny, Jakub and Sajnani, Hitesh and Vitek, Jan},
}

@inproceedings{allamanis19:_onward,
  author = {Allamanis, Miltiadis},
  title = {The Adverse Effects of Code Duplication in Machine Learning Models of Code},
  year = {2019},
  isbn = {9781450369954},
  publisher = {Association for Computing Machinery},
  address = {New York, NY, USA},
  doi = {10.1145/3359591.3359735},
  booktitle = {Proceedings of the 2019 ACM SIGPLAN International Symposium on New Ideas, New Paradigms, and Reflections on Programming and Software},
  pages = {143–153},
  numpages = {11},
  location = {Athens, Greece},
  series = {Onward! 2019}
}

@inproceedings{Zheng21,
  author={Zheng, Mingwei and Yang, Jun and Wen, Ming and Zhu, Hengcheng and Liu, Yepang and Jin, Hai},
  booktitle={2021 36th IEEE/ACM International Conference on Automated Software Engineering (ASE)},
  title={Why Do Developers Remove Lambda Expressions in {Java}?},
  year={2021},
  volume={},
  number={},
  pages={67-78},
  doi={10.1109/ASE51524.2021.9678600},
}

@article{Mazinanian17,
  author={Mazinanian, Davood and Ketkar, Ameya and Tsantalis, Nikolaos and Dig, Danny},
  title={Understanding the Use of Lambda Expressions in {Java}},
  year={2017},
  issue_date={October 2017},
  publisher={Association for Computing Machinery},
  address={New York, NY, USA},
  volume={1},
  number={OOPSLA},
  doi={10.1145/3133909},
  journal={Proc. ACM Program. Lang.},
  month={10},
  articleno={85},
  numpages={31},
}

@inproceedings{scheller2013usability,
  title={Usability evaluation of configuration-based {API} design concepts},
  author={Scheller, Thomas and K{\"u}hn, Eva},
  booktitle={International Conference on Human Factors in Computing and Informatics},
  pages={54--73},
  year={2013},
  organization={Springer}
}

\end{document}